# Milli-Tesla Quantization enabled by Tuneable Coulomb Screening in Large-Angle Twisted Graphene

I. Babich[1,2], I. Reznikov[1,2], I. Begichev[1,2], A. E. Kazantsev[3], S. Slizovskiy[3,4], D. Baranov[2], M. Siskins[2], Z. Zhan[5], P. A. Pantaleon[5], M. Trushin[1,2], J. Zhao[2], S. Grebenchuk[2], K. S. Novoselov[1,2], K. Watanabe[6], T. Taniguchi[7], V. I. Fal'ko[3,4], A. Principi[3], A. I. Berdyugin[1,8]

[1]Department of Materials Science and Engineering, National University of Singapore, Singapore, Singapore.
[2]Institute for Functional Intelligent Materials, National University of Singapore, Singapore, Singapore
[3]Department of Physics and Astronomy, University of Manchester, Manchester, UK
[4]National Graphene Institute, University of Manchester, Manchester, UK
[5]Imdea Nanoscience, Madrid, Spain
[6]Research Centre for Electronic and Optical Materials, National Institute for Material Science, Tsukuba, Japan
[7]Research Centre for Materials Nanoarchitectonics, National Institute for Material Science, Tsukuba, Japan
[8]Department of Physics, National University of Singapore, Singapore, Singapore.

**ABSTRACT**

The electronic quality of graphene has improved significantly over the past two decades, revealing novel phenomena. However, even state-of-the-art devices exhibit substantial spatial charge fluctuations originating from charged defects inside the encapsulating crystals, limiting their performance. Here, we overcome this issue by assembling devices in which graphene is encapsulated by other graphene layers while remaining electronically decoupled from them via a large twist angle (~10-30°). Doping of the encapsulating graphene layer introduces strong Coulomb screening, maximized by the sub-nanometer distance between the layers, and reduces the inhomogeneity in the adjacent layer to just a few carriers per square micrometre. The enhanced quality manifests in Landau quantization emerging at magnetic fields as low as ~5 milli-Tesla and enables resolution of a small energy gap at the Dirac point. Our encapsulation approach can be extended to other two-dimensional systems, enabling further exploration of the electronic properties of ultrapure devices.

**INTRODUCTION**

Electron mobility is a critical figure of merit for semiconductors, relevant for observation of quantum phenomena and electronic applications[1–8]. Recently, graphene was established as a material with the highest room-temperature mobility[9,10] ~150 000 cm$^2$V$^{-1}$s$^{-1}$ which is two orders of magnitude higher than that of traditional semiconductors. However, the graphene mobility at cryogenic temperatures is still lower than that of GaAs two-dimensional electron gases (2DEGs), which electronic quality has been gradually improving over many decades[4–7], and these days can reach[8] 57×10$^6$ cm$^2$V$^{-1}$s$^{-1}$ under optimal doping. In contrast, state-of-the-art graphene devices reach mobilities[9–24] of only (1-3)x10$^6$ cm$^2$V$^{-1}$s$^{-1}$, an order of magnitude lower. Potentially, this discrepancy should vanish close to the charge neutrality point (CNP), where graphene mobility theoretically diverges. However, in real devices, the electron transport near the CNP is strongly affected by the macroscopic spatial charge fluctuations[10,21–28], usually referred to as electron-hole puddles[28], limiting graphene performance.

In the first generation of devices, graphene was placed on top of Si/SiO$_2$ substrate[1–3], where it was influenced by contamination from the environment, surface roughness, and impurities, which limited mobility to $\mu \approx 10^4$ cm$^2$V$^{-1}$s$^{-1}$ and caused substantial charge density fluctuations of $\delta n \sim 10^{11}$-$10^{12}$ cm$^{-2}$. A significant improvement was achieved through the encapsulation of graphene with atomically flat hexagonal boron nitride (hBN) dielectric crystals[9–24], which reduced charge inhomogeneity to $\delta n \sim 10^{10}$ cm$^{-2}$, and improved mobility to ~$10^6$ cm$^2$V$^{-1}$s$^{-1}$. Further progress was made by using graphite gates[16–20] instead of silicon, suppressing inhomogeneity to $\delta n \sim 5 \times 10^9$ cm$^{-2}$, and establishing the current benchmark for state-of-the-art devices. Such devices enabled electron transport studies at the CNP near the room temperature, revealing properties of Dirac plasma[10,21–24]. However, at cryogenic temperatures, electron-hole puddles continue to dominate the electronic properties of graphene near the CNP. The better performance to date has been achieved in suspended graphene devices, where the dielectric substrate is eliminated, allowing charge inhomogeneity as low as[3,29–31] $\delta n \sim 4 \times 10^8$ cm$^{-2}$ and enabling to approach the Dirac point within 1 meV. However, such devices are impractical for most applications because of the challenges in fabricating dual-gated or multilayer freestanding structures.

The difference in quality between suspended and hBN-encapsulated devices is typically attributed to charged defects inside encapsulating hBN crystals, which results in charge inhomogeneity in graphene. While room for further improvement of hBN crystal quality is limited, this issue could potentially be addressed by employing Coulomb screening. Namely, a layer with a high density of states (DoS) in close proximity to graphene should suppress the electric field from charged defects and associated charge inhomogeneity.

Previous studies on Coulomb screening have utilized graphite or graphene as screening layers separated from the studied graphene by 3–10 nm thick hBN spacers[14,24,32]. While those efforts have enabled the observation of screening-induced Anderson localization[32] and suppressed particle-particle collisions[14,24,33], they did not result in a notable improvement of the device quality. As the screening depends exponentially on the distance[14], it is essential to position the screening layer much closer to the graphene than the charged defects (see Supplementary Note 6). This makes commonly used hBN spacers impractical for electron-hole puddle screening, as charged defects, which are also present inside these crystals, are located closer to graphene than the screening layer.

In this work, we address this challenge by stacking graphene layers directly atop one another and intentionally decoupling them using a large twist angle $\theta \sim 10 - 30°$. The suppressed interlayer tunneling and decoupled spectra of graphene layers in such heterostructures[34–40] result in significant interlayer resistance[34,38]. This allowed us to selectively charge one of the graphene layers in such devices and use it as a charged screening substrate separated from the test layer by a sub-nanometer vdW gap.

**RESULTS**

Fig. 1a shows one of our large angle twisted bilayer graphene (LATBG) devices with a twist angle of $\theta \sim 20°$ (See Methods for the details of the device fabrication). Using gold top and graphite bottom gates we can independently set the out-of-plane displacement field *D* and the total charge density $n_\text{tot}$.

Firstly, we characterized the LATBG device by measuring its resistance at zero *D*, as shown in Fig. 1b. Under such conditions, both layers have the same doping, and the system behaves similarly to single-layer graphene: resistance sharply peaks at zero carrier density and rapidly drops when doping increases. Using this curve, we estimate inhomogeneity of individual layers $\delta n \sim 7 \times 10^9 \text{cm}^{-2}$, and electron mobility $\mu_e = 0.5 \times 10^6$ cm$^2$/Vs (Supplementary Fig. 1), which is consistent with that of a typical high-quality encapsulated devices reported in the literature[9–24].

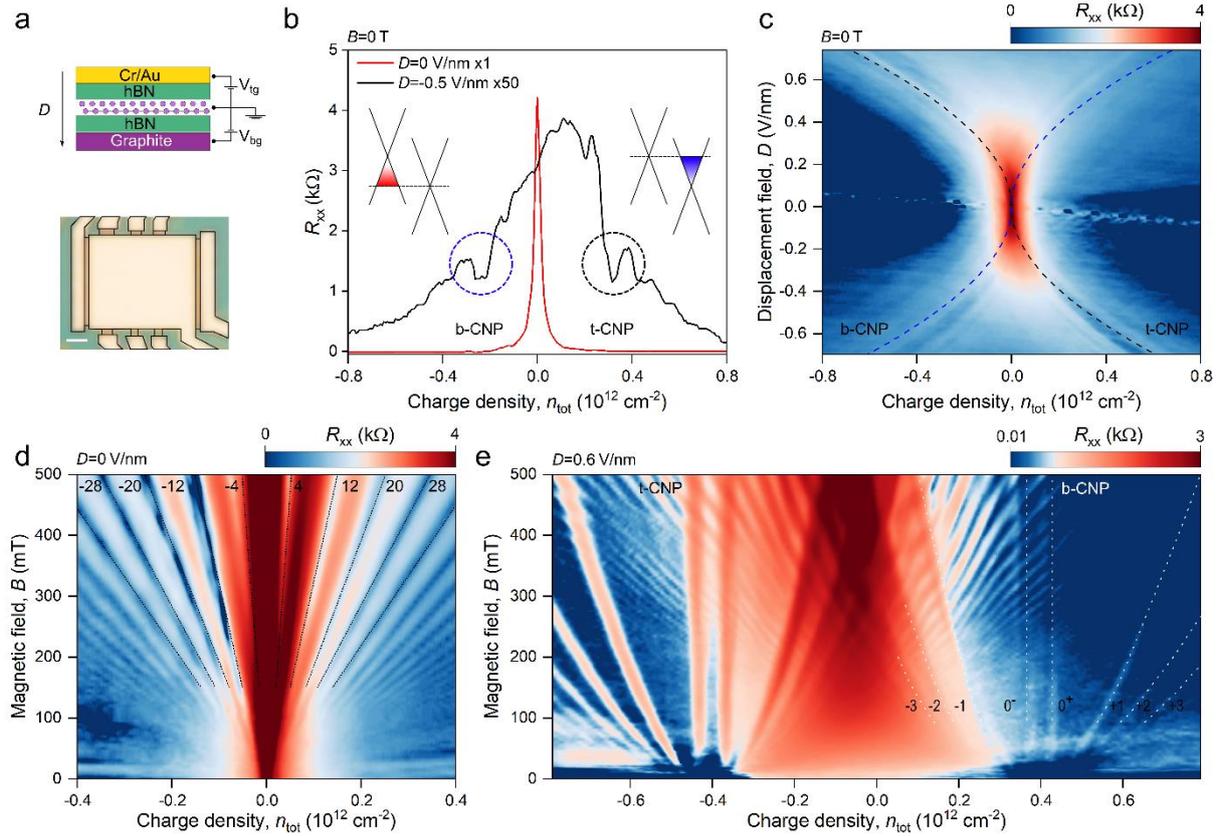

**Figure 1. Characterization of a large-angle twisted bilayer graphene (LATBG) device. a**, Top panel: schematic structure of LATBG device, utilizing thin hBN flakes as dielectrics for bottom graphite and top metal gates, $V_{tg}$ and $V_{bg}$ are top and bottom gate voltages; bottom panel: optical image of LATBG Hall bar device. Black line highlights the metallic top gate and electrical contacts. Scale bar is 1 $\mu$m. **b**, Longitudinal resistivity as a function of total charge density $n_{tot}$ for *D*=0 V/nm, and *D*=0.5 V/nm measured at zero magnetic field. Dashed circles indicate the positions of charge neutrality points of top (t-CNP, black circle) and bottom graphene layers (b-CNP, blue circle) calculated using the electrostatic model (see Supplementary Note 1). Insets schematically show band structure of LATBG under applied D when the Fermi level crosses the CNP of one of the graphene layers; here, the red part of the cone represents hole doping, and blue represents electron doping of graphene. **c**, Longitudinal resistivity at B=0 T as a function of $n_{tot}$ and D. Dashed lines indicate expected positions of CNPs. **d, e**, Resistivity as a function of magnetic field B and $n_{tot}$ for D=0 V/nm (**d**) and D=0.6 V/nm (**e**). In panel **d**, dashed lines show the expected position of doubled graphene filling factors. In panel **e**, white dashed lines are a guide for the eye of the first three landau levels and CNP gap boundaries in bottom graphene layer. See the text for the further discussion. Further examples of Landau fan measurements are shown in Supplementary Figs. 2, 4. The high magnetoresistance observed around $n_{tot} = 0$ can be attributed to the compensated semimetal state, similar to Ref.[24], which naturally forms under applied D when one layer is electron-doped and the other is hole-doped. Measurements were performed at 2 K for all panels.

Qualitatively, the applied *D* creates an interlayer potential difference that separates the Dirac cones in energy, whereas $n_{tot}$ moves the common Fermi level (see inset Fig. 1b). When the Fermi level crosses the Dirac points, it is reflected as resistivity wiggles at positive and negative doping levels, as shown in Fig. 1b. The dual-gate map in Fig. 1c further shows how Dirac cone offset evolves upon changing *D*. Dashed lines mark the expected positions of Dirac points of top and bottom layers which coincide well with resistivity features that we attributed to CNPs earlier.

**Screening enabled quantization in milli-Tesla magnetic field**

Next, to test the device quality we measured Landau fan diagrams under different displacement fields. When *D*=0 V/nm (Fig. 1d) LATBG shows a typical Landau fan diagram of single-layer graphene, but with doubled filling factors, as expected for two graphene layers with equal doping. The applied $D$ significantly alters this picture: in Fig. 1e, there are two sets of fan diagrams converging around $n_{\text{tot}} = \pm 0.4 \times 10^{12}$ cm$^{-2}$, which correspond to the expected positions of the top and bottom graphene CNPs. These fans can be attributed to the individual quantization of the top and bottom layers, and their parabolic-like shape originates from the presence of the other heavily doped layer (see the schematic band structure in Fig. 2a and Supplementary Note 1). If plotted as a function of the charge density in top or bottom layer, the fan diagrams restore their linearity (Supplementary Fig. 2) and become similar to those shown in Fig. 1d.

An important difference between the fan diagrams observed in Figs. 1d and 1e is the magnetic field required to resolve the onset of Landau quantization. At $D$=0 V/nm Landau levels (LLs) become resolvable around $B^* \approx 100$ mT, while the applied displacement field significantly lowers this onset. To determine the onset of oscillations under an applied $D$ we have zoomed into a small magnetic field range, as shown in Fig. 2b. In this map, the signatures of Landau fans become visible already at $B^* = 5 - 6$ mT (see Supplementary Fig. 11 and Supplementary Note 5), which is an order of magnitude better than the magnetic field required to see the Landau quantisation at zero $D$ in LATBG and in test graphene devices without proximity screening (Supplementary Fig. 8).

A reduction of magnetic field required to resolve the onset of quantization in Figs. 1e and 2b indicates a decrease of charge inhomogeneity $\delta n$ under applied $D$. Qualitatively, to resolve the first cyclotron gap in graphene, the fluctuations of the Fermi level must be smaller than the size of the first cyclotron gap[30]. To crosscheck this criterion, we modelled DoS of graphene for given inhomogeneity level, magnetic field and temperature as a function of Fermi level and energy fluctuations $\delta E$ (see Supplementary Fig. 9 and Supplementary Note 4). When $D$=0 V/nm and $B^* = 100$ mT ($B^*$ is the smallest magnetic field allowing resolution of the first LL), the model suggests $\delta n \approx 5 \times 10^9$ cm$^{-2}$ which agrees with earlier estimations. Under applied $D$ (for $B^* = 5 - 6$ mT) we find that inhomogeneity drops to $\delta n \approx 2 - 3 \times 10^8$ cm$^{-2}$, corresponding to just 2-3 electrons per micrometre area of the Hall bar shown in Fig. 1a. However, at small magnetic fields, the cyclotron radius $R_c$ becomes comparable to the width of our voltage probes $W$, which possibly limits the resolution of quantization onset. To show this, in Fig. 2c, we plotted the $W$=2$R_c$ condition, which well describes the onset of Landau quantization in our device, indicating that the quality of the LATBG device likely to be even better than estimated above. Finally, while quantization becomes apparent already at 5–6 mT, we estimate that the quantum Hall effect fully onsets at 13 mT (see Supplementary Note 5). However, at low $B$, zero resistivity within the cyclotron gaps is not observed because of parallel conduction through the second strongly doped layer, which remains non-quantized and contributes a significant background signal.

The suppression of $\delta n$ under applied displacement field originates from the tuneable screening, in agreement with the design of our experiment. The screening of an external electric field by metallic layers is set by the DoS around the Fermi energy, which drops to zero when both layers are simultaneously tuned towards CNP ($D$=0 case). As a result, any charged impurities within the encapsulating hBN crystals create substantial spatial fluctuations in carrier density, as illustrated in the top panel of Fig. 2c. On the contrary, the applied $D$ creates an energy offset between Dirac cones of different layers which entails that the double layer device has high DoS even when one of graphene layers is tuned towards the CNP. It results in screening of external electric field and improves the homogeneity of graphene layers, as shown in bottom panel of Fig. 2c. This behavior can be quantitatively described using the Thomas-Fermi model (see Supplementary Note 6) and 2D maps in Fig. 2d which show how the screening layer suppresses charge inhomogeneity. We also note that the analysis above primarily focuses on the energy gap between the 0th and 1st Landau levels. This gap is the largest and, due to the low intrinsic doping of graphene at this filling, is expected to be the most sensitive to external screening. In contrast, higher Landau level gaps typically emerge only at larger carrier densities, where enhanced self-screening within the quantized graphene layer reduces its susceptibility to external screening.

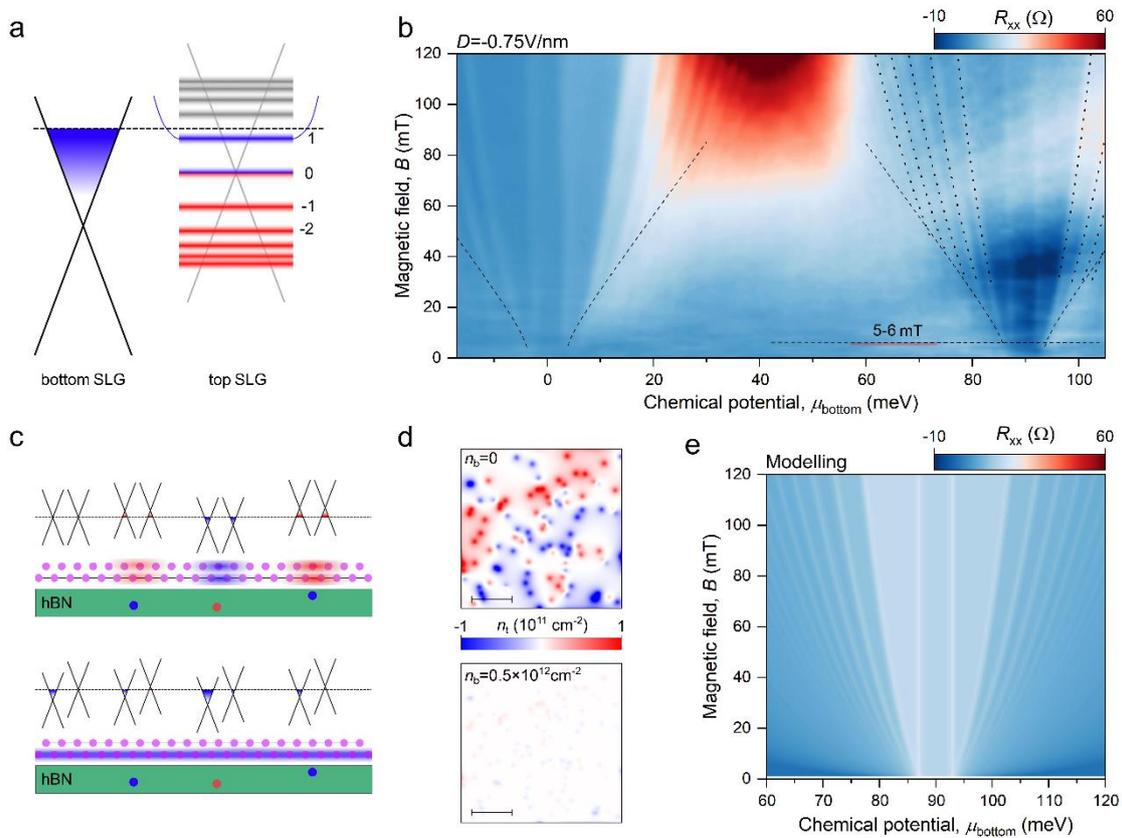

**Figure 2. Resolving the onset of Landau quantization in milli-Tesla magnetic field**. **a**, Schematic illustration of LATBG band structure at low magnetic field $B$ and high $D$ when Fermi level is tuned close to the CNP of one of graphene layers. **b**, Fan diagram measured at $D$=-0.75 V/nm and 2 K shown as a function of chemical potential in the bottom graphene layer $\mu_b$. Black parabolic dashed lines indicate the expected position for the first five Landau levels (LLs) plotted using a standard graphene sequence $E_N = \sqrt{2\hbar e B v_F^2 N + \Delta^2}$, where $\hbar$ is the reduced Plank constant, $e$ is the electron charge, $E_N$ is the energy of the $N^{th}$ LL, and $v_F$ is the Fermi velocity in graphene, $2\Delta$ is a band gap discussed further in the text. The horizontal dashed line and the error bar mark the onset of Landau quantisation. Square root dashed lines indicate the limit of quantization set by the probe width (see Supplementary Note 5). **c**, Schematic illustrations of electron-hole puddles in LATBG at zero (top) and under applied (bottom) displacement fields. Red and blue shaded regions represent positive and negative doping correspondingly, coloured circles in the hBN illustrate charged defects. **d**, Simulated charge density profiles in top graphene calculated for hBN with impurity density $n_{imp} = 10^{10} \text{cm}^{-2}$ at zero doping of the bottom layer $n_b = 0$, and for $n_b = 0.5 \times 10^{12} \text{cm}^{-2}$ (see Supplementary Note 6). Scale bars are 250 nm. **e**, Modelling of LATBG resistance subjected to high $D$ under the assumption of a gapped graphene spectra (see Supplementary Note 7).

## Resolving the gap at the CNP of graphene layers

Another notable difference between the measurements at zero and the applied $D$ is the presence of two magnetic field-independent resistance peaks in the centres of the fan diagrams of each graphene layer in Figs. 1e and 2b. To understand these features, we perform spectroscopy of the top layer by plotting its fan diagram as a function of chemical potential in the bottom layer $\mu_b$, see Fig. 2b. The theoretical positions of the first few LLs, indicated by dashed lines, align well with the resistivity features on this map, except that instead of a single vertical resistivity peak expected for zero-energy LL, there are two peaks. This suggests the formation of a gap at the CNPs, estimated to be $\Delta_g = 5 \pm 1$ meV in both layers. Furthermore, we found that the gap is present even at zero magnetic field, as shown in high-resolution fan diagram, magnetic focusing measurements, and bulk-current fan diagram (Supplementary Figs. 2, 4). It is independent of $D$ and becomes resolvable already at $D$=0.05V/nm (see Supplementary Figs. 3, 5).

To verify the presence of the band gap, we modelled the fan diagram of a double-layer graphene system, assuming a gap at the CNP of the quantized layer. Usually, such a gap should produce a local resistance maximum, which is not observed in our devices. To reproduce the measurements shown in Fig. 2b, we had to additionally assume valley decoupling in the gapped graphene layer, as discussed in Supplementary Note 7, which leads to the edge conductivity channels inside the gap. The results of our modelling Fig. 2e capture the measurements shown in Fig. 2b, confirming that the two vertical resistance peaks originate from the gap edges, with their separation set by the gap size. The possible origin of this gap is discussed further in the text.

**Graphene fully encapsulated with screening graphene layers**

Apart from LATBG devices, where graphene serves as a substrate for another graphene layer, we fabricated large-angle twisted trilayer graphene (LATTG) devices to showcase full encapsulation of the middle graphene layer. Optical images of fabricated devices are shown in Fig. 3a. See Methods for fabrication details.

To characterize LATTG, we firstly measured resistance as a function of $D$ and $n_{tot}$, as shown in Fig. 3b. Similar to bilayer devices, the resistance map reveals features that evolve under an applied displacement field. Using an extended electrostatic model (see Supplementary Note 1), we identified these features as the CNPs of the individual graphene layers, indicated by dashed lines in Fig. 3b. While symmetry considerations suggest that the middle-layer CNP (m-CNP) should remain independent of $D$, the m-CNP in Fig. 3b shifts towards negative doping with increasing $D$. This behavior is reproducible across all our devices and can be explained by small energy offsets of outer graphene layers interfacing hBN[39].

Next, we measured the same $D$ vs. $n_{tot}$ map at a small magnetic field of $B$=50 mT, as shown in Fig. 3c. At this field, all three layers become quantized and produce three sets of parallel resistivity lines (labeled by corresponding LL number). There are single resistivity peaks corresponding to zeroth LLs in top and middle layer, while in the bottom layer, the peak doubles (labeled as 0+ and 0–). This feature, similarly to observations in LATBG, corresponds to the formation of a gap. However, in this LATTG device, the gap is resolved only in one of the layers.

To further assess the quality of our LATTG devices, we measured fan diagrams under fixed $D$ or $n_{tot}$, shown in Fig. 3d-g and Supplementary Fig. 22. In all LATTG devices, we observed three sets of Landau fans converging at carrier densities corresponding to the expected positions of the CNPs for individual layers, as indicated by the arrows in Figs. 3d and 3f. Notably, in Fig. 3f, the complete lifting of the zeroth Landau level degeneracy is resolvable at magnetic fields as low as 1 Tesla, while it is not seen in devices without screening at similar field ranges. The high-resolution fan diagrams at small magnetic fields shown in Figs. 3e and 3g reveal the onset of Landau quantization already at 5–7 mT (see Supplementary Note 5 and Supplementary Fig. 12) for two different LATTG devices, showcasing the high electronic quality of graphene in such structures.

**DISCUSSION**

While both LATBG and LATTG devices exhibit similar onsets of quantization in millitesla magnetic field range, LATTG devices demonstrate superior electronic quality. For instance, at $B$=50 mT LATTG device allows to resolve up to 20-30 LLs per layer, Fig. 3c, whereas LATBG devices show only up to 8 LLs even at higher $B$=100 mT, Supplementary Fig. 5. However, in LATTG devices, when the Fermi level is tuned towards the CNP of one of the layers, the other two heavily doped layers act as parallel conduction channels. This obstructs the access to the properties of the charge neutral layer by introducing significant background signals that can interfere with measurements (see Supplementary Fig. 22). Another difference in performance between LATBG and LATTG devices is that the former shows identical gaps at CNPs of both layers (see Fig. 2d), whereas the latter shows a clear gap only in one of the layers (see Fig. 3c, and Supplementary Fig. 22). Below, we consider several possibilities for the gap origin.

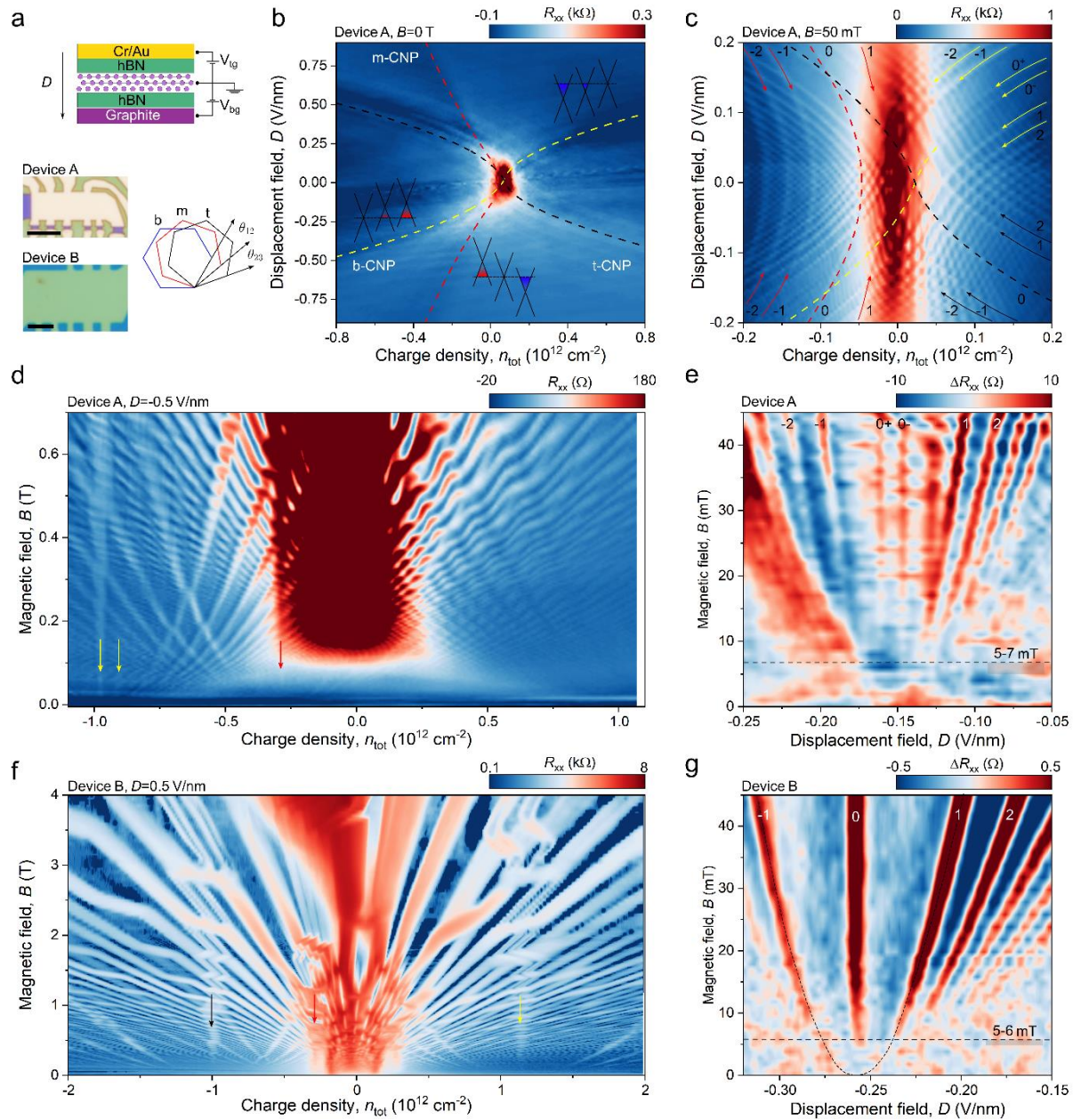

**Figure 3. High quality of LATTG devices. a,** Schematic structure of LATTG devices, graphene layers (b, m and t labels indicate bottom, middle and top layers respectively) are twisted by large angles $\theta_{12}$ (the angle between the bottom and the middle layers) and $\theta_{23}$ (the angle between the middle and the top layers). See Methods for further information. Optical images show two measured LATTG Hall bar devices A and B, scale bar is 5 μm. **b,** Longitudinal resistance at $B=0$ T as a function of charge density $n_{tot}$ and displacement field $D$ measured in device A. Coloured lines indicate conditions of CNP for top, middle and bottom layers with black, red, and yellow correspondingly. Inset band structures illustrate the position of CNPs of each layer. **c,** Longitudinal resistance at $B=50$ mT as a function of $n_{tot}$ and $D$ measured in device A. Coloured dashed lines show the calculated CNP positions for all three layers based on the electrostatic model described in the Supplementary Note 1. We therefore label the nearest Landau levels as ±1, ±2, with the sign reflecting the charge carrier type in each layer.

**d-g,** Magnetoresistance measurements for devices A and B. (**d, f**), measured for fixed *D*=-0.5 V/nm and *D*=0.5 V/nm correspondingly as a function of $n_{\text{tot}}$, coloured arrows show positions of CNPs. (**e, g**) longitudinal resistance measured for fixed $n_{\text{tot}}$ as a function of *D* with removed background (see Supplementary Fig. 10 for more details). Onset of Landau quantisation become resolvable already at $B^* = 5 - 7$ mT in (**e**) and at $B^* = 5 - 6$ mT in (**g**). Black dashed parabolic curve is a guide for the eye of the ±1 Landau levels. Measurements were done at 2 K for all panels.

Multiple theoretical works have predicted a Mott insulator gap at the CNP of graphene[41,42]. However, experiments with freestanding graphene have shown renormalization of the Fermi velocity instead of a gap, which was attributed to the presence of strong long-range Coulomb interactions[29]. A recent study suggests that suppressing the tail of Coulomb interactions (e.g., with a metallic screening layer) may result in a semimetal-Mott insulator transition[42]. However, our modelling of such transition predicts gap dependence on temperature and *D*, which are not observed in the measured devices (see Supplementary Notes 2, 8 and Supplementary Figs. 5,6). This points towards a structural origin of the gap, which may appear due to the contact between graphene and hBN (G/hBN) or due to the twist between graphene layers.

Prior studies on the devices where graphene is aligned with encapsulating hBN crystals report varying gap sizes depending on the graphene-hBN alignment[17,18,43–48]; however this effect is minimized in our samples, since we intentionally misaligned graphene and hBN during device fabrication. Moreover, in Fig. 2d, the observed gaps in the LATBG device are identical in both layers; this symmetry would likely be broken if misaligned hBN/G interface had a noticeable impact on the band structure. Next, in LATTG devices, the presence of a gap in only one of the outer layers suggests that any gap due to the graphene/hBN interface can indeed be negligible with our intentional misalignment (otherwise, it would be resolved in the other outer layer) pointing to the possible intrinsic origin of this gap within the twisted graphene structure.

To investigate this hypothesis, we calculated the band structure of LATBG at various high commensurate angles (Supplementary Fig. 23). The calculated band structures revealed a small band gap of a few meV opening at the Dirac points of each layer, along with a larger gap at the CNP of the double layer, which was not observed experimentally. It is essential to note that these calculations are influenced by the choice of the unit cell origin, which may change due to the lateral shift between the layers (see Supplementary Note 10).

It should be noted that existing theories on the band structure of LATBG mainly focus on commensurate angles[49–51] which are challenging to achieve in real devices. Moreover, the precise determination of the electronic structures is highly sensitive to the actual arrangement of atoms, as determined by the lateral shift and twist angle. These challenges make the systematic study of gaps at the Dirac points quite difficult and place it beyond the scope of this work.

Although it is challenging to pinpoint the exact microscopic origin of the gap, the ability to resolve such small features of the band structure highlights the spectroscopic capabilities enabled by encapsulation with graphene layers. Since resolving gaps below 5 meV ($\delta E < 2.5$ meV) at the CNP of Dirac point requires a doping inhomogeneity level below $2\times10^9$ cm$^{-2}$, our samples exceed performance of the state-of-the-art devices. More than that, tunable Coulomb screening reduces doping inhomogeneities towards $3\times10^8$ cm$^{-2}$, which is an order of magnitude improvement and corresponds to $\delta E \approx 0.5\ meV$. Such high quality allowed to resolve the onset of Landau quantization in our devices at magnetic fields of 5-6 mT, corresponding to a quantum mobility relevant for the observation of quantum phenomena, $\mu_q \approx (1.5\text{-}2)\times10^6$ cm$^2$V$^{-1}$s$^{-1}$. Unfortunately, the large-angle twisted multilayer graphene geometry does not allow the direct measurement of the transport mobility of individual graphene layers under applied screening, as a significant portion of the current always propagates through a heavily doped graphene layer. Nevertheless, our toy model (see Supplementary Note 3 and Supplementary Fig. 7) estimates that the transport mobility near CNP in our devices approaches $\mu \approx$ (12-20)x$10^6$ cm$^2$V$^{-1}$s$^{-1}$, an order of magnitude higher than $\mu_q$. This difference is expected as transport mobility is limited by backscattering, whereas quantum mobility is more sensitive to small-angle scattering. Both quantum and transport mobilities in our devices are an order of magnitude higher than those in devices without screening. We also note that the quantum mobility achieved in our devices is higher than the quantum mobility of the best GaAs 2DEGs, where quantum oscillations onset[7] around 35-40 mT even at milli-Kelvin temperature, with the quantum mobility[6] $\mu_q \approx$ 1x$10^6$ cm$^2$V$^{-1}$s$^{-1}$. At the same time, our estimations give mobility values of the same order as that of the best GaAs samples, highlighting a significant improvement of graphene devices.

Finally, we note that the large twist angle approach also introduces new capabilities, such as tuneable Coulomb interactions via adjustable screening, facilitating further exploration of many-body effects in moiré quantum materials[52]. Our encapsulation method is readily adaptable to other multilayer and twisted heterostructures, thereby broadening its potential applications.

**METHODS**

**Sample fabrication**

The heterostructures studied in this work were assembled using a combination of the tear-and-stack method together with dry transfer technique[9,12,53]. Polydimethylsiloxane (PDMS) stamps coated with a thin polycarbonate (PC) membrane were employed to pick up exfoliated crystals from Si/SiO$_2$ substrates. The pick-up process was conducted at 100 °C in the following sequence: hexagonal boron nitride (hBN) of 30–40 nm thickness (top layer), a large graphene flake (first segment), the second segment of the same graphene flake, rotated to a large twist angle, and another hBN layer (bottom layer, 30–40 nm). Our target angle for the HATBG sample is 20°, for the HATTG devices A and B $\theta_{12} \approx 20°, \theta_{23} \approx -20°$ and $\theta_{12} \approx 30°, \theta_{23} \approx 30°$ respectively; the device C shown in the Supplementary Information had target angles $\theta_{12} \approx 10°, \theta_{23} \approx 10°$. The final vdW stack was released on top of graphite flake at 180 °, after which the PDMS stamp was carefully delaminated from the PC membrane. The melted PC film was subsequently dissolved in dichloromethane, followed by rinsing in acetone and isopropyl alcohol.

Next, we used high-resolution atomic force microscopy (AFM) to locate bubble-free regions suitable for device fabrication. Standard nanofabrication techniques were then used to pattern Hall bars in the identified clean regions. First, the top gates were fabricated using electron-beam lithography (EBL), followed by chromium/gold (Cr/Au) metal deposition. In a subsequent EBL step, deep reactive ion etching (DRIE) with a CHF$_3$/O$_2$ gas mixture was employed to define the Hall bar geometry. Finally, one-dimensional electrical contacts to the twisted bilayer graphene (TBG) were formed by depositing Cr/Au onto the exposed graphene edges.

**Electronic transport measurements**

The electronic transport measurements were performed using standard low-frequency lock-in techniques with excitation currents below 100 nA, minimizing heating and non-linear effects; the measurement temperature was 2K if not specified. Most of the data presented in the main text were acquired under nonzero displacement field, *D*. The dual-gated device configuration allowed independent control of *D* and the total carrier density, $n_{\text{tot}}$. The carrier density was determined as the sum of the densities induced by the top and bottom gates: $n_{\text{tot}} = \frac{1}{e}(C_{\text{tg}}V_{\text{tg}} + C_{\text{bg}}V_{\text{bg}})$, where $V_{\text{tg}}$ and

$V_{\text{bg}}$ are the voltages applied on a top and bottom gates, and $C_{\text{tg}}$ and $C_{\text{bg}}$ are the top and bottom gate capacitances per unit area, respectively, found from the Hall effect measurements. The displacement field was calculated as: $D = \frac{1}{2\varepsilon_0}(C_{\text{tg}}V_{\text{tg}} - C_{\text{bg}}V_{\text{bg}})$, where $\varepsilon_0$ is the vacuum permittivity.

To characterize our devices, we first measured their resistivity at zero displacement and magnetic field and defined inhomogeneity as a half-width at half-maximum (HWHM) as illustrated in Supplementary Fig. 1a. We note that the inhomogeneity of individual layers is half of the HWHM. Next, we calculated mean free path and mobility using a standard expression for the single layer graphene but considering that we have two graphene sheets instead of one:

$$\mu = \frac{1}{\rho n_{\text{tot}} e}; \tag{1}$$

$$l_{\text{mfp}} = \frac{1}{\rho}\frac{\hbar}{e^2}\sqrt{\frac{\pi}{2n_{\text{tot}}}}. \tag{2}$$

Here $\rho$ is the resistivity of the whole double layer, $\hbar$ is reduced Plank constant, and $e$ is an electron charge. The results are shown in Supplementary Fig. 1b,c, which indicate that at zero $D$ the mean free path in our device is approaching device width, and mobility approaching $10^6$ cm$^2$V$^{-1}$s$^{-1}$ as expected for the high-quality encapsulated devices.

At applied $D$, we cannot decipher resistivity of the individual layer, however, magnetic focusing measurements allow to claim when the sample is ballistic. In Supplementary Fig. 2, magnetic focusing line approaches the boundaries of the gap as close as the measurement accuracy $\delta n < 2 \times 10^9\ cm^{-2}$, and corresponds to the mean free path of $l_{mfp} > L = 2.7$μm, where $L$ is the distance between centers of the contacts, and the inequality sign is used because the electrons move along cyclotron orbits. Thus, we can set a lower boundary for electron mobility at $D$=-0.55 V/nm:

$$\mu = \frac{l_{\text{mfp}} e}{\hbar}\sqrt{\frac{2}{\pi \delta n}} > 7.3 \times 10^6\ \text{cm}^2/\text{Vs}. \tag{3}$$

**DATA AVALIABILITY**
Relevant data supporting the key findings of this study are available within the article and the Supplementary Information file. All raw data generated during the current study are available from the corresponding authors upon request.

# Supplementary Information for 'Milli-Tesla Quantization enabled by Tuneable Coulomb Screening in Large-Angle Twisted Graphene

I. Babich[1,2], I. Reznikov[1,2], I. Begichev[1,2], A. E. Kazantsev[3], S. Slizovskiy[3,4], D. Baranov[2], M. Siskins[2], Z. Zhan[5], P. A. Pantaleon[5], M. Trushin[1,2], J. Zhao[2], S. Grebenchuk[2], K. S. Novoselov[1,2], K. Watanabe[6], T. Taniguchi[6], V. I. Fal'ko[3,4], A. Principi[3], A. I. Berdyugin[1,7]

[1]Department of Materials Science and Engineering, National University of Singapore, Singapore, Singapore.
[2]Institute for Functional Intelligent Materials, National University of Singapore, Singapore, Singapore
[3]Department of Physics and Astronomy, University of Manchester, Manchester, UK
[4]National Graphene Institute, University of Manchester, Manchester, UK
[5]Imdea Nanoscience, Madrid, Spain
[6]National Institute for Materials Science, Tsukuba, Japan
[7]Department of Physics, National University of Singapore, Singapore, Singapore.

**List of Contents.**
Supplementary Figures 1-22

Supplementary Tables 1, 2

Supplementary Note 1. Calculating layer specific charge density in LATBG and LATTG devices.
Supplementary Note 2. Additional characterization of the band gap observed at the CNP of graphene.
Supplementary Note 3. Estimation of transport electron mobility in our devices.
Supplementary Note 4. Onset of Landau quantization in single layer graphene devices.
Supplementary Note 5. Determination of quantization onset.
Supplementary Note 6. Charge inhomogeneity screening model.
Supplementary Note 7. LATBG fan diagram modelling.
Supplementary Note 8. Possibility of a many-body gap at CNP of graphene
Supplementary Note 9. Additional LATTG device
Supplementary Note 10. LATBG bandsturcture calculations

Supplementary References

## Supplementary Figures

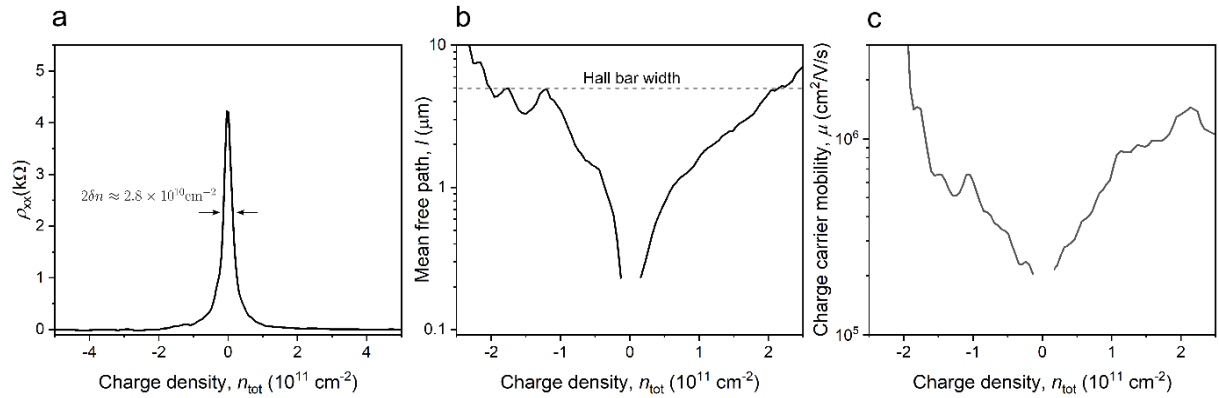

**Supplementary Figure 1. a**, resistivity as a function of carrier density measured at zero $D$ at $T=2$K for the device shown in Fig. 1a of the main text. **b**, mean free path calculated from the resistivity shown in the panel **a**. **c**, the mobility of LATBG obtained using data from the panel **a**.

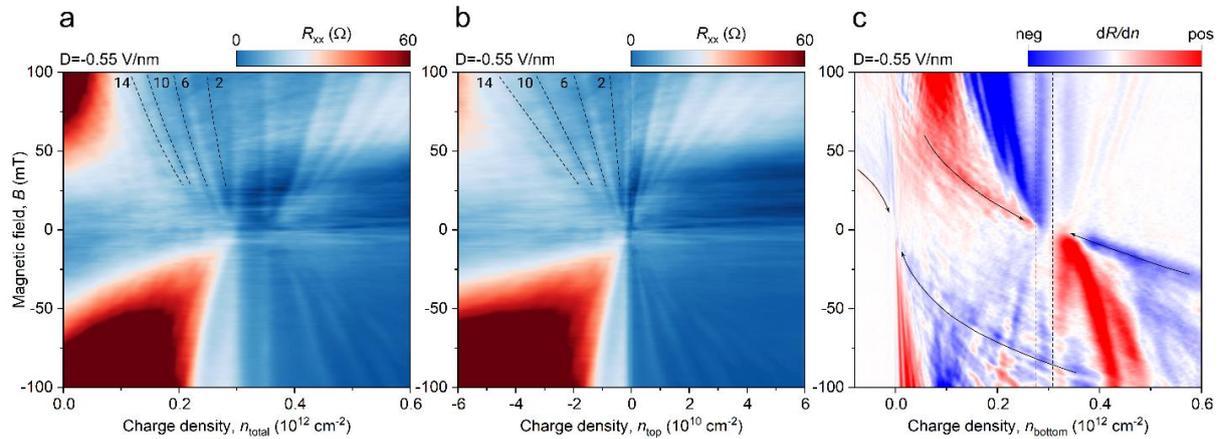

**Supplementary Figure 2. a**, Fan diagram of the main device under applied $D=-0.55$ V/nm. Dashed lines are a guide for an eye for the first four fully filled Landau levels (LLs). **b**, Fan diagram from (**a**) plotted as a function of charge density in top graphene layer. Parabolic-like LLs restore linear behavior: dashed lines indicate conditions of full filling of first four LLs from (**a**). **c**, First derivative of resistance as a function of magnetic field and charge density of the bottom graphene layer, measured in a magnetic focusing geometry. Black arrow lines are a guide for an eye for magnetic focusing peaks which converge towards gap boundaries.

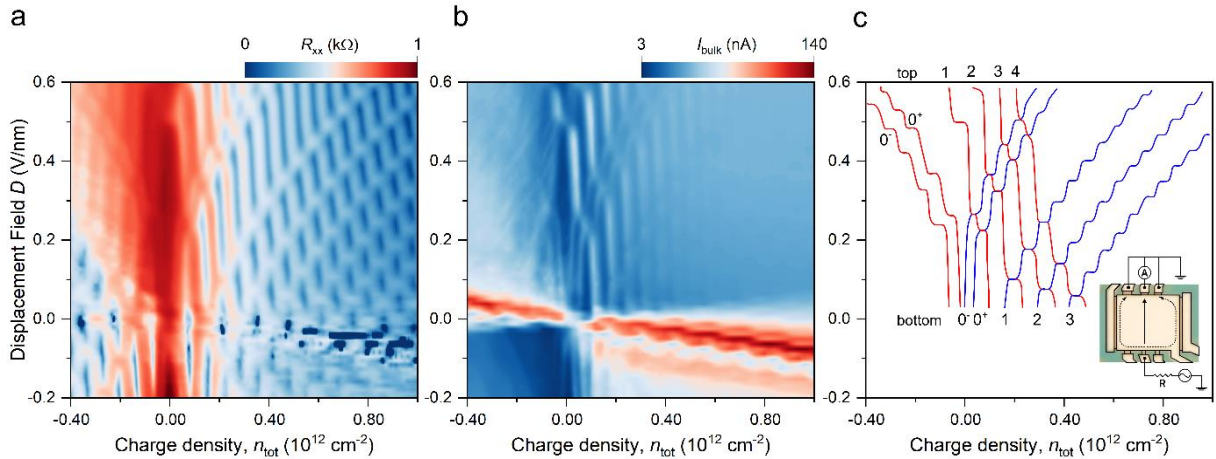

**Supplementary Figure 3**. **a**, Resistance as a function of $D$ and $n_{tot}$. Peaks in resistance correspond to half-filled Landau levels (LLs), whereas dips correspond to crossing of cyclotron gaps in both layers. **b**, Bulk current as a function of displacement field and $n_{tot}$. Peaks in the current correspond to half-filled LLs, whereas dips indicate insulating state in the cyclotron gaps. **c**, Schematic illustration of LL structure in **a** and **b**. Blue lines indicate LLs in the bottom layer, including boundaries of the gap at CNP (labelled as $0^+$, $0^-$) red lines indicate LLs in the top layer. The inset shows measurement geometry used for the panel b. Such geometry is sensitive only to the current going through the bulk of the sample, cutting away the edge channels[5].

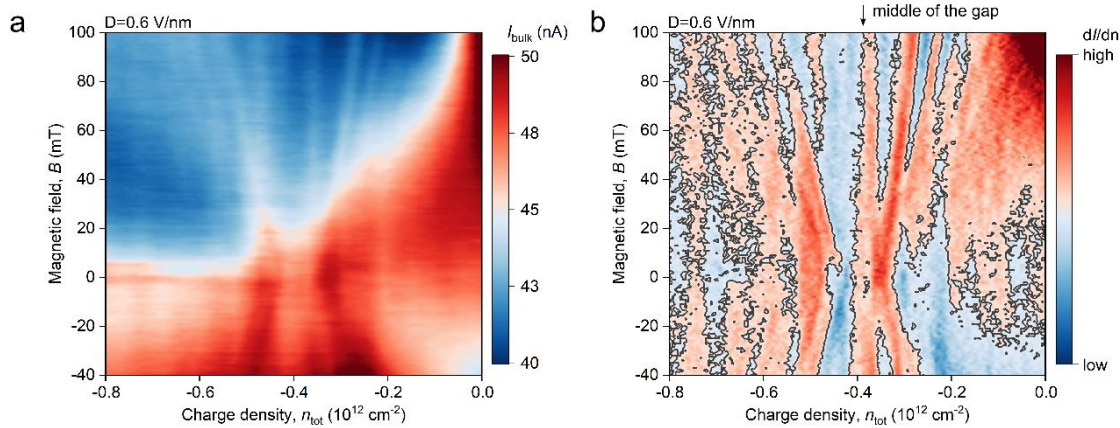

**Supplementary Figure 4. a**, Fan diagram measured at D=0.6 V/nm in a measurement geometry from Ref 5, which allows to eliminate contribution of the edge channels. Higher values of bulk current reflect better bulk conductance. **b**, Derivative of the map in **a**, the black contour lines indicate zero derivative condition to highlight that resistivity maximum at the bulk is present in any magnetic field.

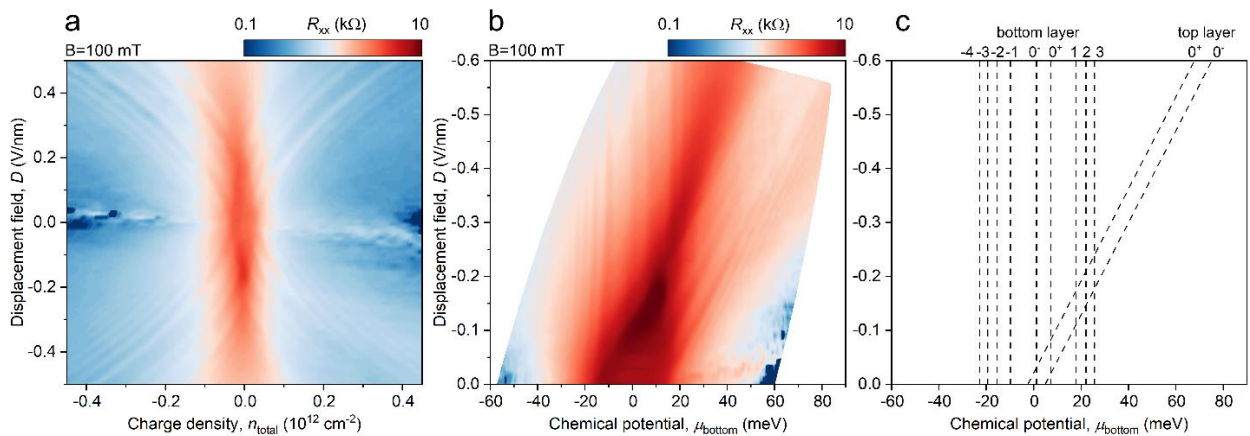

**Supplementary Figure 5. a**, Dual-gate map of resistance at 100 mT as a function of total charge density and displacement field. **b**, Data from (**a**) plotted as a function of chemical potential of bottom graphene layer and displacement field. **c**,

Transcription of Landau levels (LLs) from (**a**) and (**b**). Sequences of LLs in both layers remain independent of *D*, including the gap at half-filling.

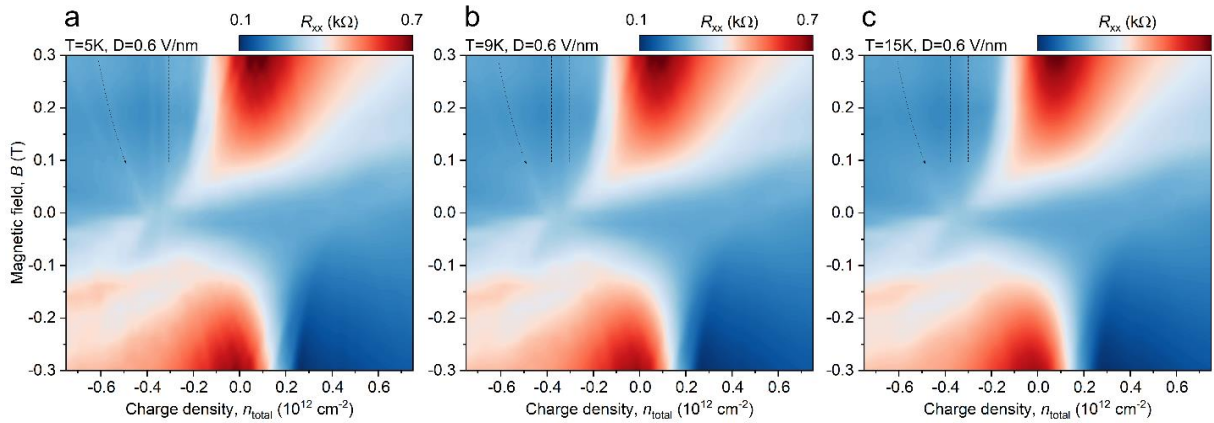

**Supplementary Figure 6. a-c**, Show fan diagrams under applied *D*=0.6 V/nm for temperatures *T*=5, 9 and 15 respectively. Dashed lines indicate gap boundaries and half-filling of first Landau level. Notably, the gap size remains the same under varying temperature, and onset of quantization remains as low as 20mT at 15K.

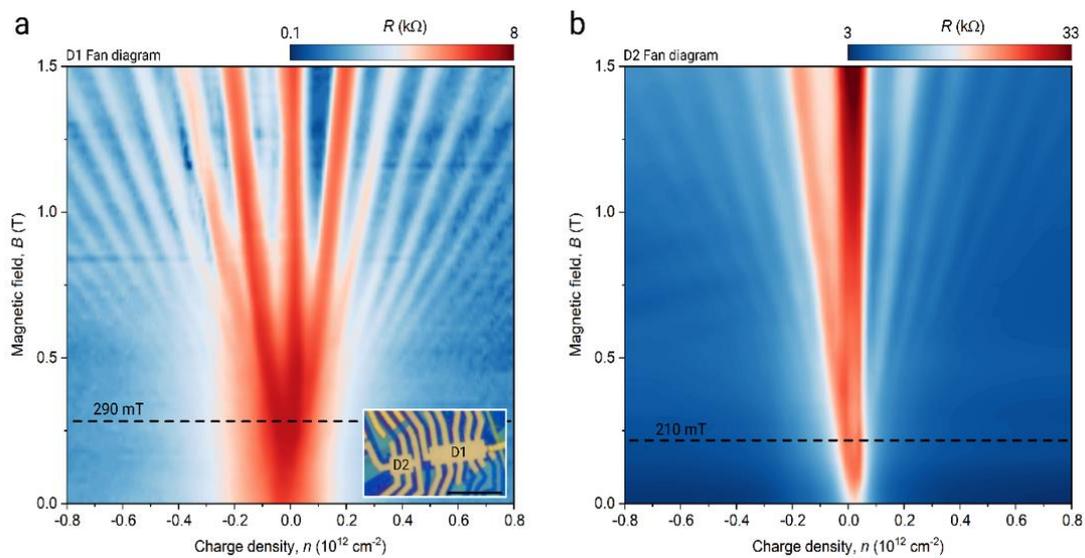

**Supplementary Figure 8. a**, resistance as a function of charge density *n* and magnetic field *B* of device D1. Dashed line at $B^* = 290$ mT level indicates onset of Landau quantization. Inset image shows graphite-gated SLG devices D1 and D2, scalebar is 10 μm. **b**, Two-probe measurements of resistance as a function of *n* and *B* for device D2. Dashed line indicates the onset of quantization at 210 mT.

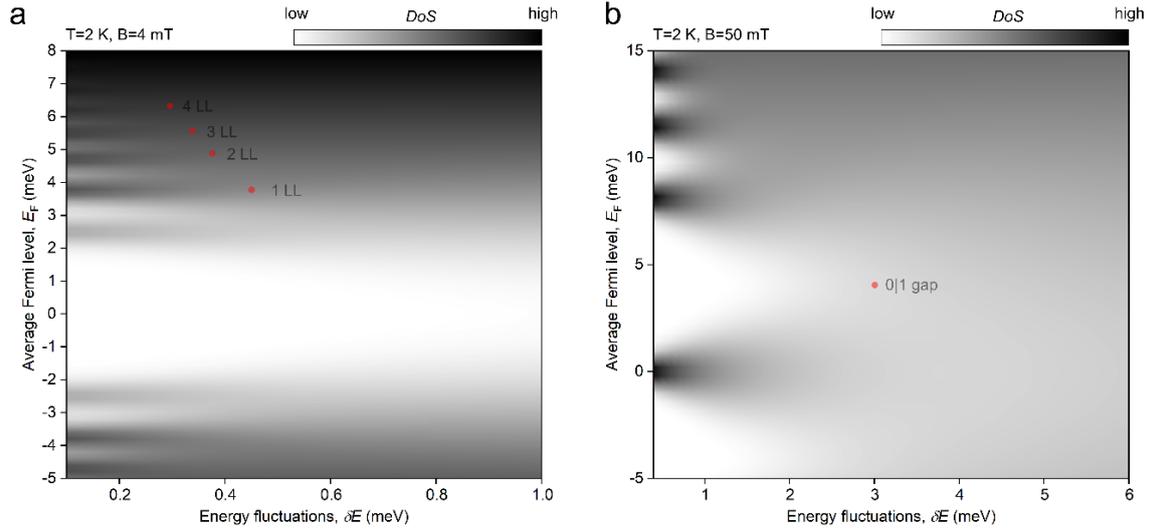

**Supplementary Figure 9. a**, Heatmap of *DoS* as a function of average position of Fermi level $E_F$ in the sample, and level of energy fluctuations $\delta E$, calculated at *T*=2K, and *B*=6mT, assuming a gap at the CNP of graphene. Dark regions correspond to Landau levels, and bright regions to cyclotron gaps. Red circles indicate onset of corresponding LL resolution. **b**, The same heatmap as in (**a**) but for 50 mT, DoS contrast of the first cyclotron gap completely disappears at 3 meV.

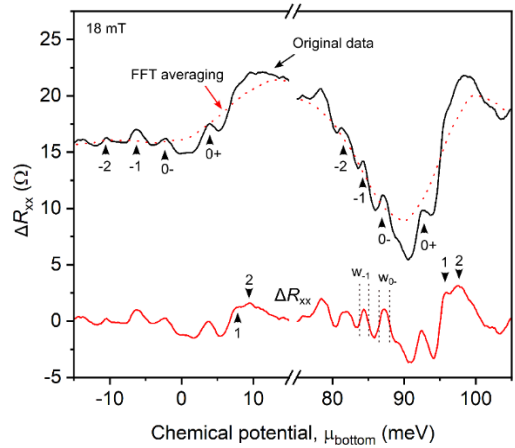

**Supplementary Figure 10.** This figure shows resistance curves from LATBG device (Fig. 2b of main text) at 18 mT. The thick black line and dotted red line are original data R and FFT smoothed data $\langle R \rangle$, whereas the solid red line is $\Delta R = R - \langle R \rangle$. Arrow tips indicate resistance peaks corresponding to LLs in top and bottom graphene. Dashed vertical lines show the FWHM of N=-1 and 0⁻ LL.

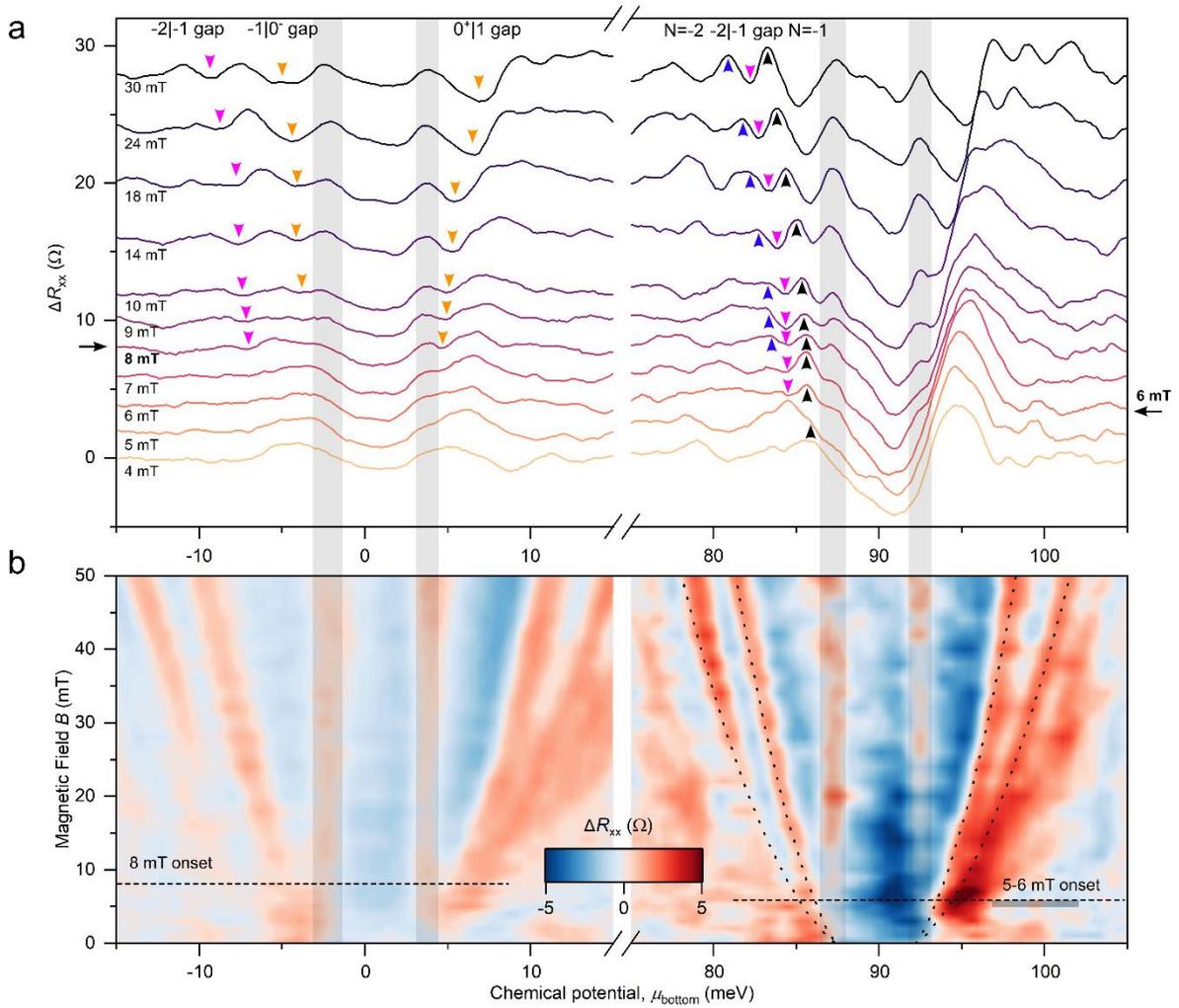

**Supplementary Figure 11. a**, Resistance curves with a subtracted background signal for fixed values of magnetic field. The arrow tips point at the resistance peaks and dips associated with LLs and energy gaps between LLs for the first few well-pronounced sequences. Arrows on the sides indicate the smallest magnetic field, at which LL peak or LL gap dip magnitude is larger than that of the background signal. The vertical grey stripes represent the edges of the zeroth LL. **b**, Magnetoresistance measurements with subtracted background. The parabolic dashed lines show the LL fit for the first two LLs. The horizontal dashed lines correspond to the onset values found in (**a**), and the grey error bar shows quantization onset range combined with remnant magnetic field determination error.

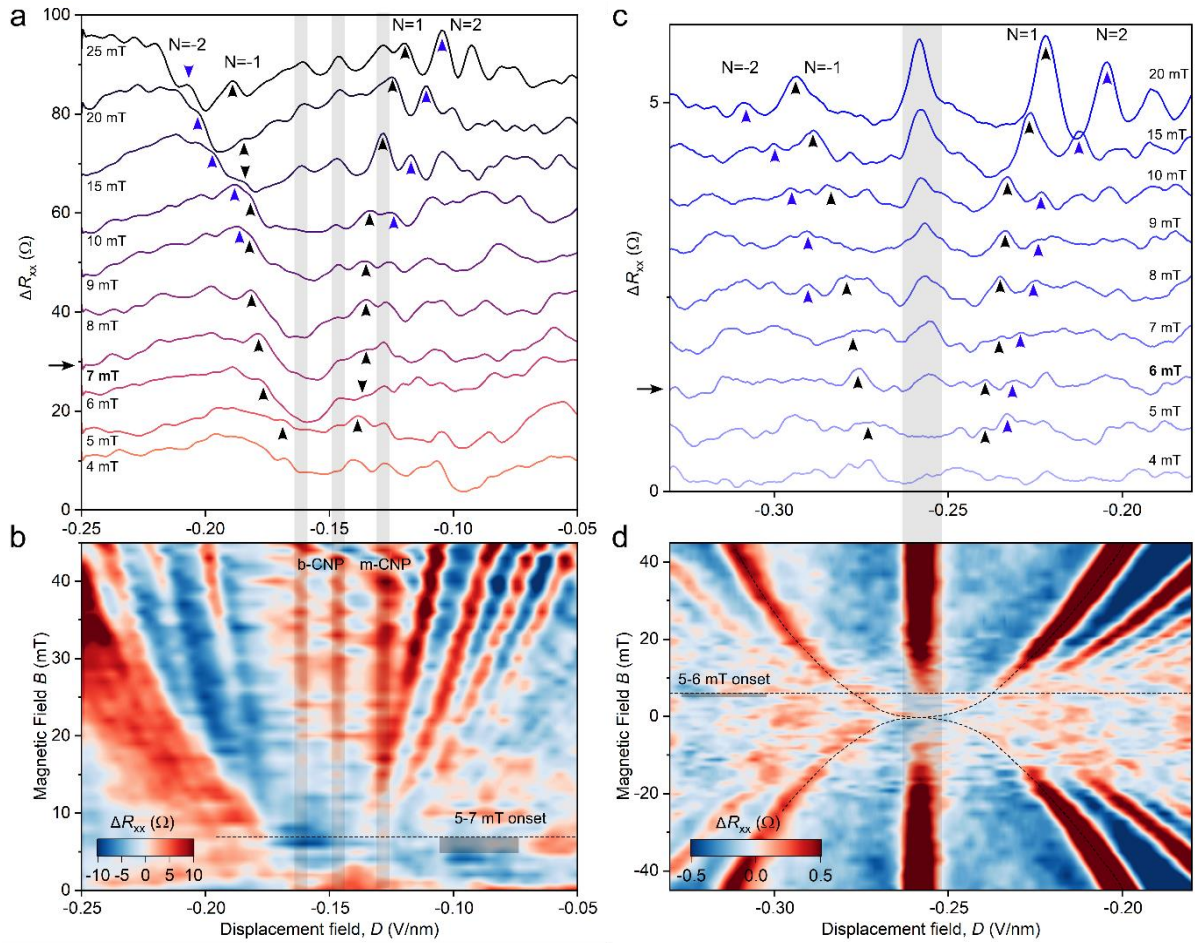

**Supplementary Figure 12. a,** Resistance curves with a subtracted background signal for fixed values of magnetic field for LATTG device A. The arrow tips point at resistance peaks associated with LLs for the first few well-pronounced sequences. Arrow on the sides indicate the smallest magnetic field, at which LL peak magnitude is larger than that of the background signal. The vertical grey stripes represent zeroth LL features in the bottom and middle layers. **b,** Magnetoresistance measurements with subtracted background for device A. The horizontal dashed line corresponds to the onset values found in (**a**), and the grey error bar shows quantization onset range combined with remnant magnetic field determination error. **c** and **d,** The same as (**a,b**) but for the device B. The dashed parabolic lines are a guide for an eye to show expected positions of the ±1 LL peak.

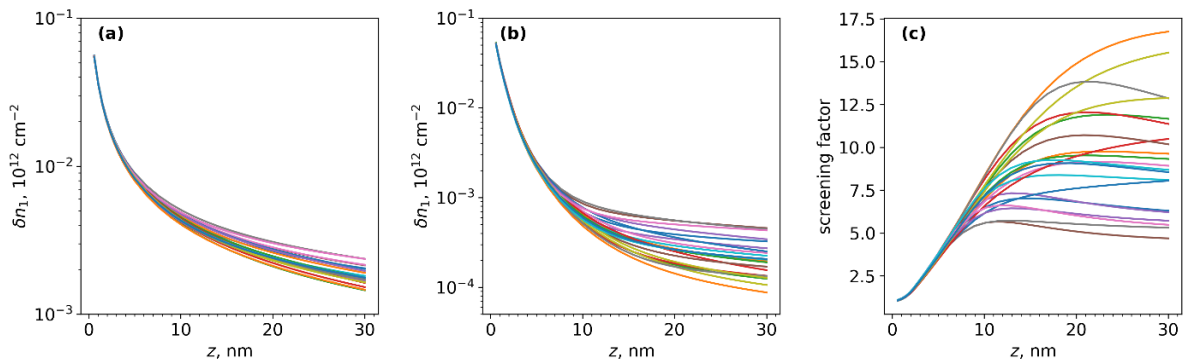

**Supplementary Figure 13. a,** density inhomogeneity in layer 1 as function of distance to the impurities at zero doping of the screening layer $\bar{n}_2 = 0 \ cm^{-2}$ and $d = 0.33 \ nm$. **b**, density inhomogeneity in layer 1 as function of distance to the impurities at $\bar{n}_2 = 0.5 \cdot 10^{12} cm^{-2}$ and $d = 0.33 \ nm$. **c**, ratio of density inhomogeneity of Panel (**a**) and that of Panel (**b**) referred to as screening factor. Different curves correspond to different disorder realizations.

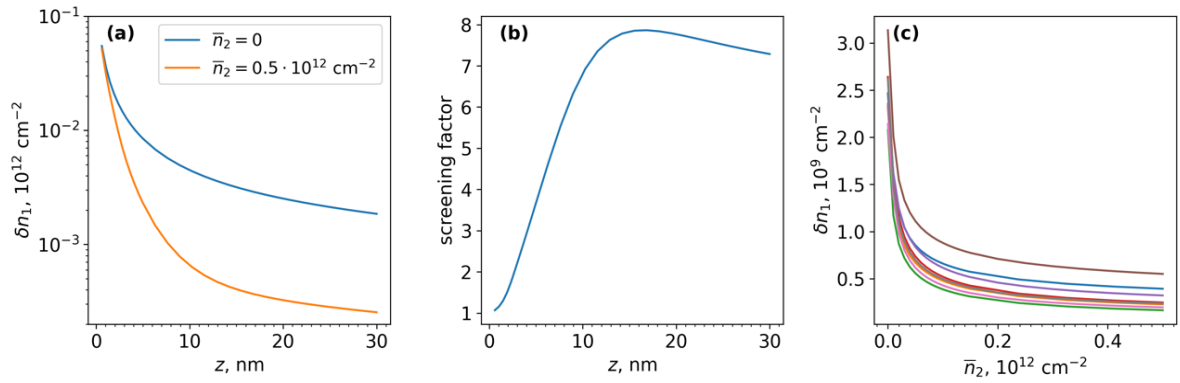

**Supplementary Figure 14**. **a**, average inhomogeneity as function of distance to impurities at $d = 0.33\ nm$ and at $\bar{n}_2 = 0\ cm^{-2}$ (blue line) and at $\bar{n}_2 = 0.5 \cdot 10^{12}\ cm^{-2}$ (red line). **b**, ratio of the two inhomogeneities in Panel (**a**). **c**, Inhomogeneity as a function of carrier density of a screening layer for different realizations of impurity configurations. Impurities are located on the same side as the screening layer.

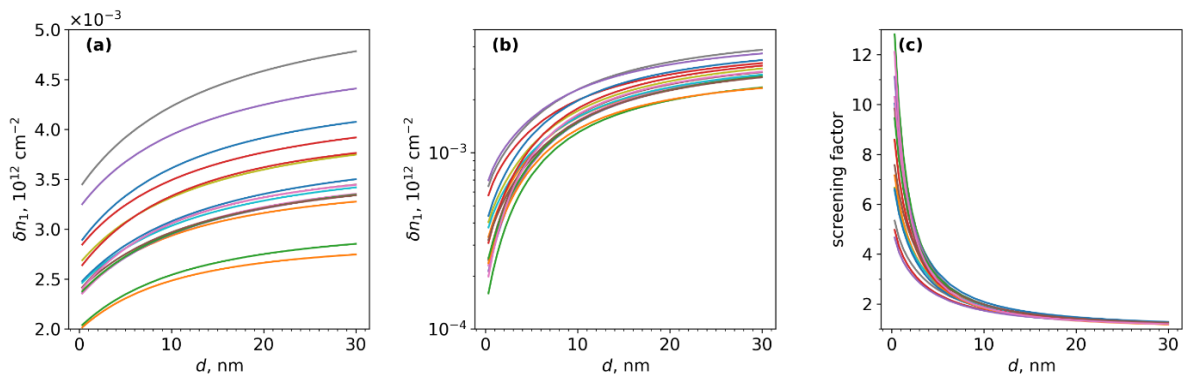

**Supplementary Figure 15**. **a**, density inhomogeneity in layer 1 as function of inter graphene layer distance at $\bar{n}_2 = 0\ cm^{-2}$ and $|z| = 20\ nm$. **b**, Density inhomogeneity in layer 1 as function of interlayer distance at $\bar{n}_2 = 0.5 \cdot 10^{12}\ cm^{-2}$ and $|z| = 20\ nm$. **c**, ratio of density inhomogeneities in Panel (**a**) and Panel (**b**). Different curves correspond to different disorder realizations, 17 in total.

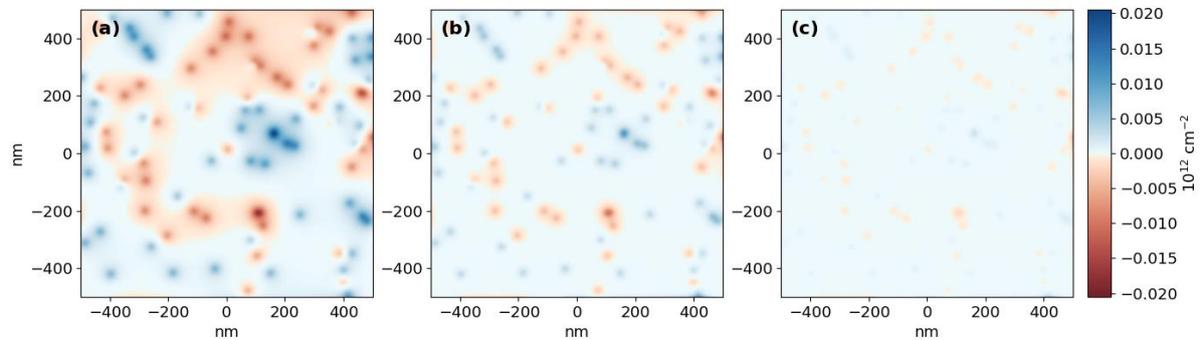

**Supplementary Figure 16**. **a**, density fluctuations in graphene layer 1 with layer 2 neutral ($d$ does not affect the figure). **b**, Density fluctuations in layer 1 with layer 2 charged to $\bar{n}_2 = 0.5 \cdot 10^{12}\ cm^{-2}$ and placed at a distance $d = 5nm$. **c**, density fluctuations in layer 1 with layer 2 charged to $\bar{n}_2 = 0.5 \cdot 10^{12}\ cm^{-2}$ and placed at $d = 0.33\ nm$. In all panels impurities with average density $n_{imp} = 10^{10}\ cm^{-2}$ are placed at a distance $|z| = 20\ nm$ away from layer 1.

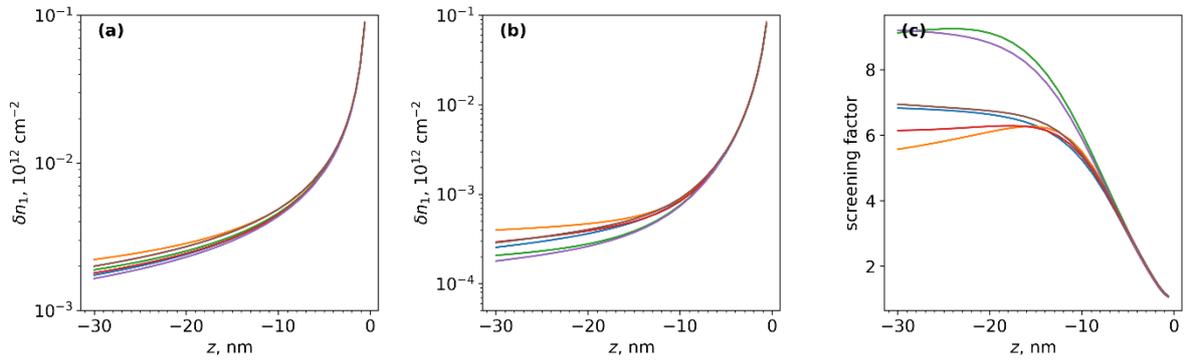

**Supplementary Figure 17. a,** density inhomogeneity in layer 1 as function of distance to the impurities at $\bar{n}_2 = 0\ cm^{-2}$ and $d = 0.33\ nm$. **b**, density inhomogeneity in layer 1 as function of distance to the impurities at $\bar{n}_2 = 0.5 \cdot 10^{12} cm^{-2}$ and $d = 0.33\ nm$. **c**, ratio of density inhomogeneity of Panel (**a**) and that of Panel (**b**) referred to as screening factor. Different curves correspond to different disorder realizations. Impurities are located on the same side as layer 1.

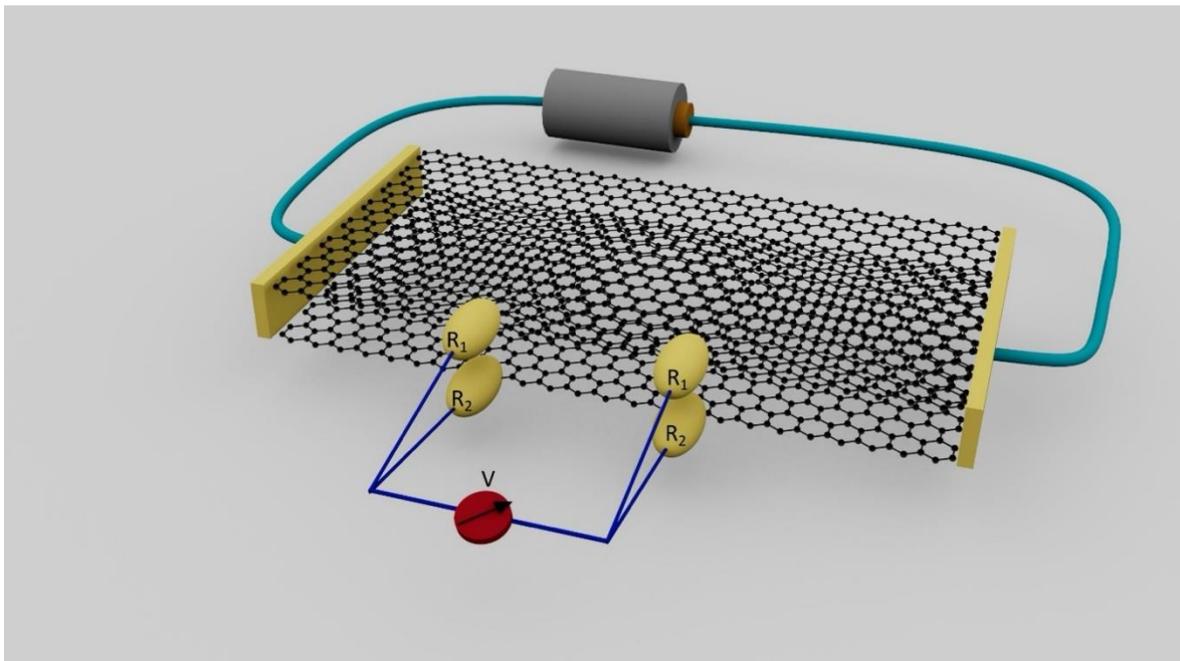

**Supplementary Figure 18.** Schematic illustration of 4-terminal geometry, showing a finite resistance of side contacts.

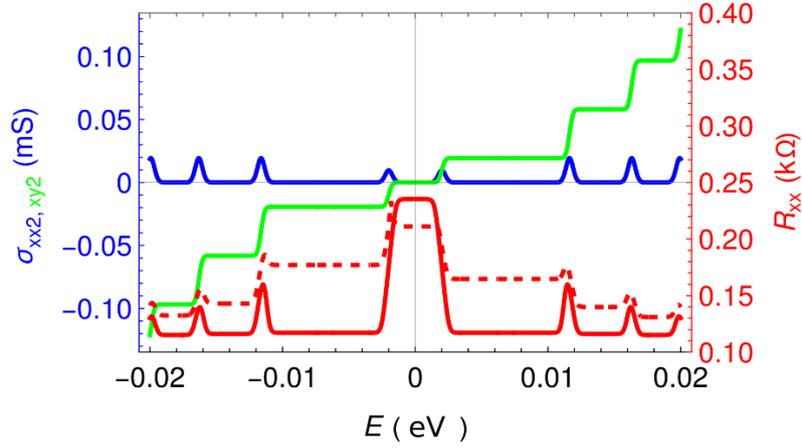

**Supplementary Figure 19**. Example of longitudinal and transversal conductivities near charge neutrality point of graphene and $R_{xx}$ calculated according to eq (28). Here, B=0.1 T and a 4 meV gap is assumed. (We use the parameters $\sigma_{xx1} = 4\,mS$; $\sigma_{xy1} = 1mS$; $a = 1$; $\sigma_{c1} = \sigma_{c2}$). We show $R_{xx}$ for zero (solid-red) and small non-zero (dashed red) tunnelling between the layers.

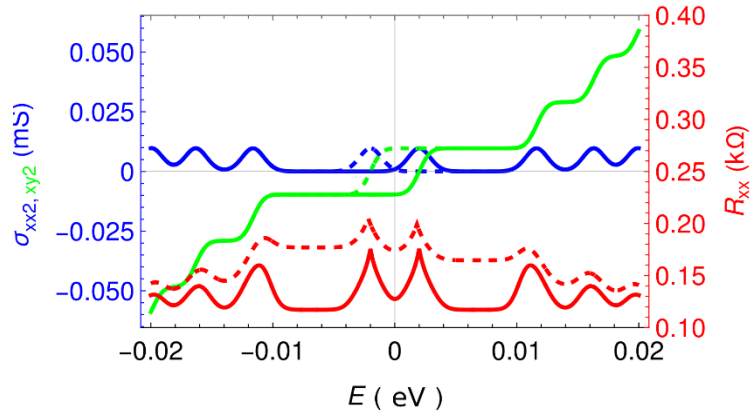

**Supplementary Figure 20**. Example of longitudinal and transversal conductivities near charge neutrality point of graphene with decoupled valleys (shown with solid/dashed lines blue and green lines) and $R_{xx}$ calculated according to eq (31) with and without small tunneling. Here, $B$=0.1 T and a 4 meV gap is assumed. (We use the parameters $\sigma_{xx1} = 4$ mS; $\sigma_{xy1} = 1$mS; $a = 1$; $\sigma_{c1} = \sigma_{c2}$). We show $R_{xx}$ for zero (solid-red) and small non-zero (dashed red) tunnelling between the layers.

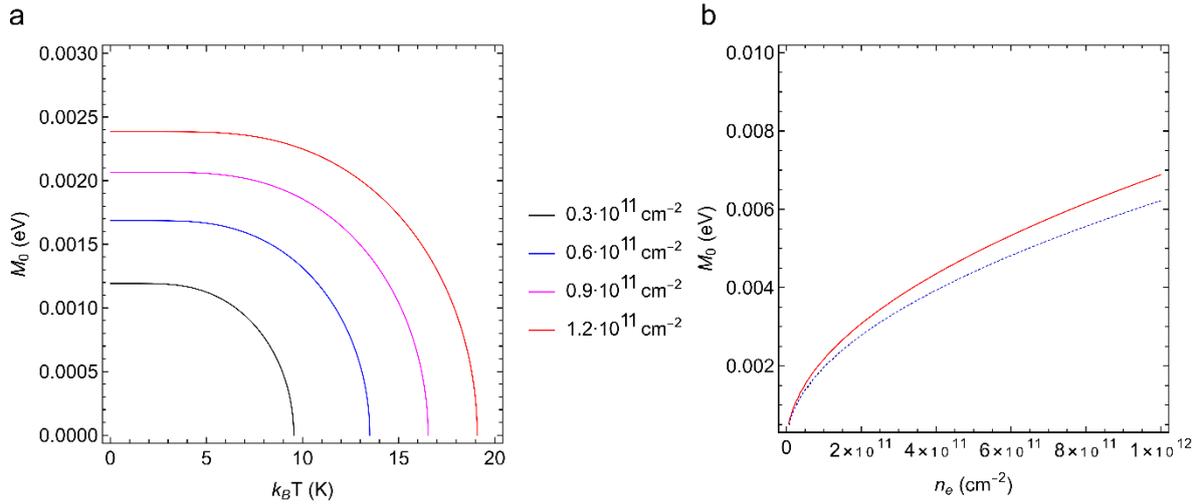

**Supplementary Figure 21. a**, Mott insulator gap calculated for fixed $n_e$. **b**, Red curve shows gap dependence on doping of the screening layer, calculated for $\varepsilon^* = 6.85$ at $T$=0. Blue dashed line shows the analytical approximation.

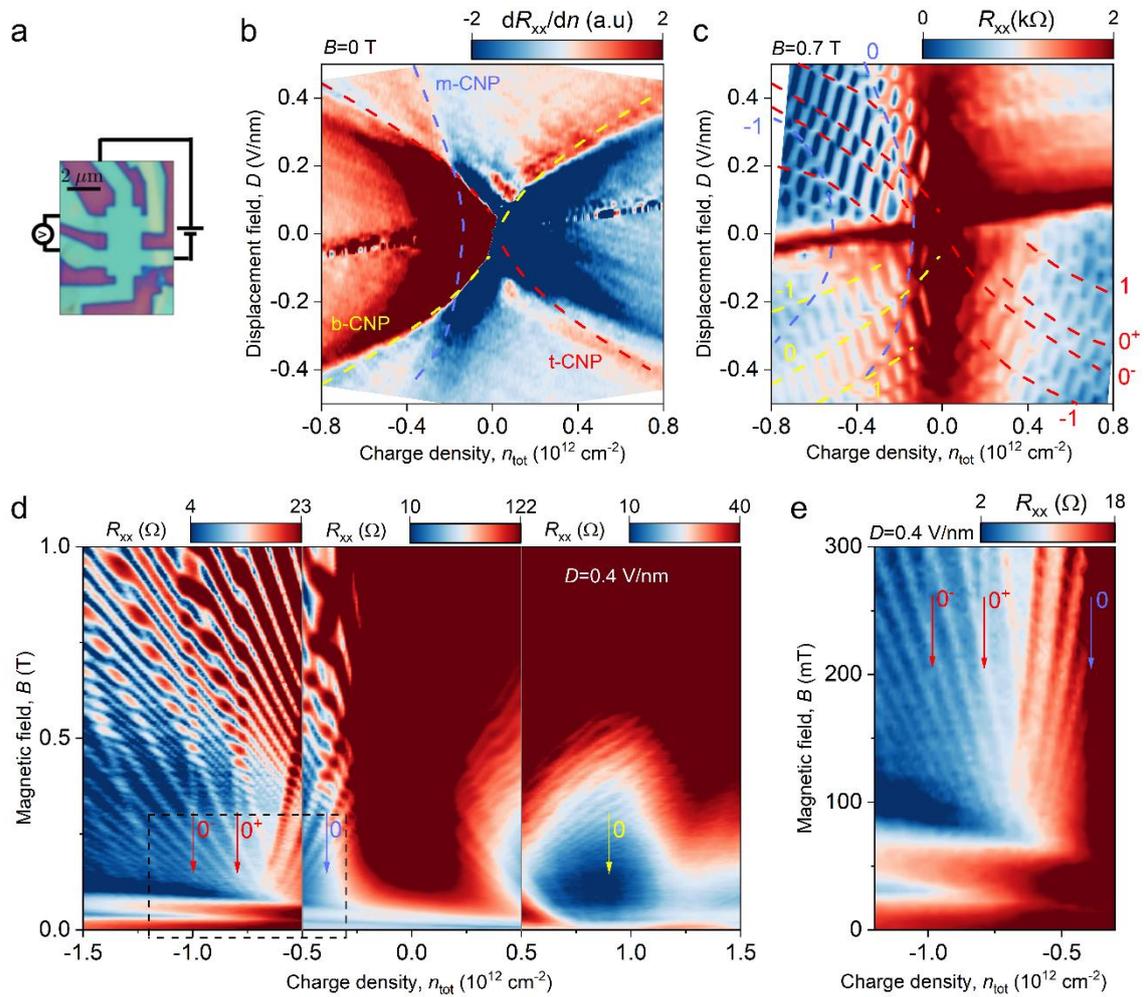

**Supplementary Figure 22. a,** Optical image of additional LATTG Hall bar device. Scale bar, 2 $\mu$m. **b,** Resistance at $B$=0 T as a function of charge density $n_{tot}$ and displacement field $D$. Red, yellow and blue dashed lines correspond to the charge neutrality conditions in the top, bottom, and middle graphene layers, respectively. **c,** Resistance at $B$=0.7 T as a function of $n_{tot}$ and $D$. Labels indicate Landau level index. **d, e,** Magnetoresistance measurements at $D$=0.4 V/nm. Red arrows highlight gap boundaries for the CNP in the top layer, which is estimated as $9 \pm 1$ meV in the same way as for LATBG. **e,** Low magnetic field range measurements

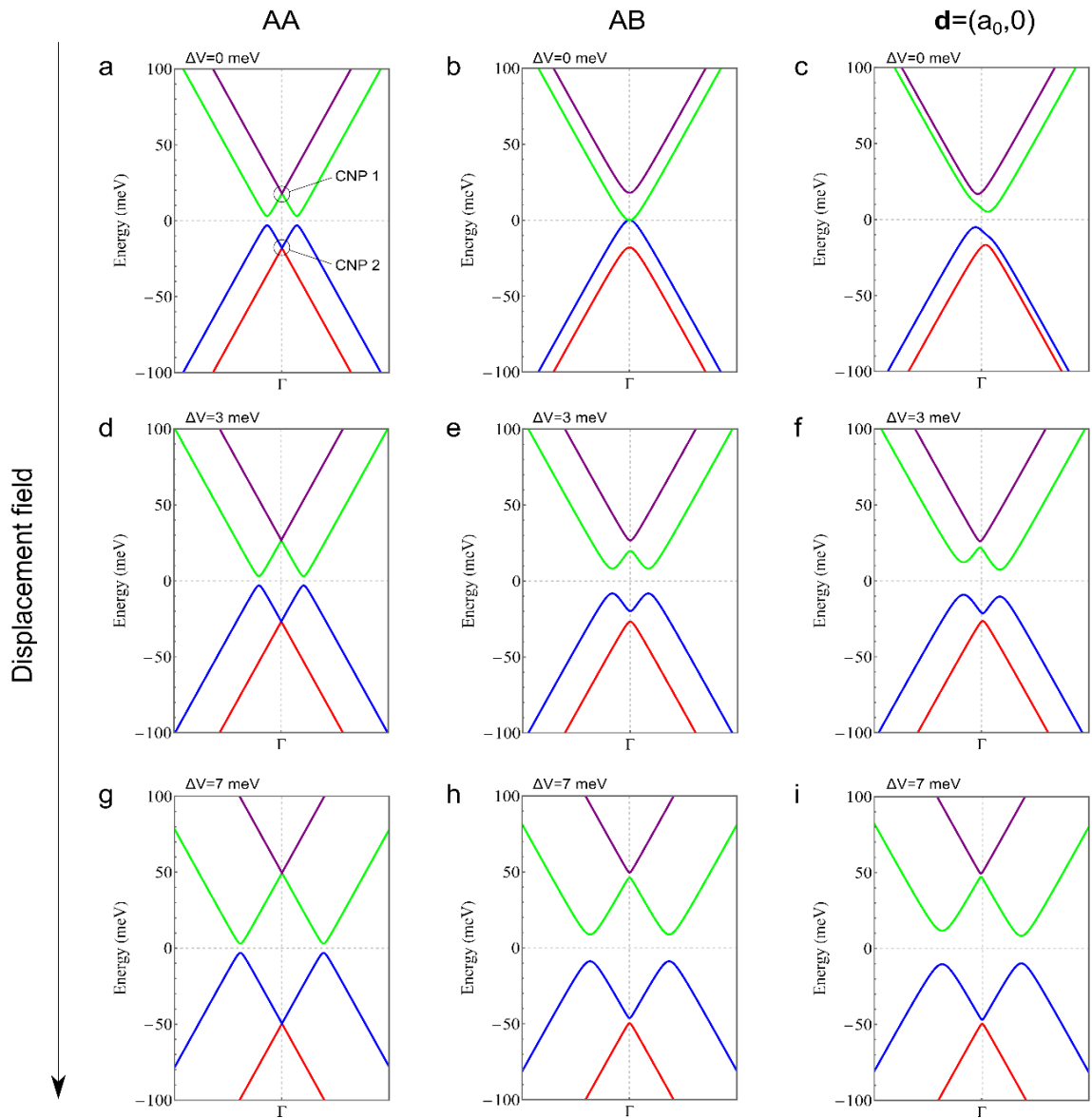

**Supplementary Figure 23.** low-energy bands of LATBG obtained using the continuum model (54) for the twist angle 21.8°. **a-c**, the evolution of band structure for different stacking configurations between the layers at zero interlayer potential $V = -$ meV; **d-f**, the same as (**a-c**) but for an interlayer potential of $V = 3$ meV; **g-i**, the same as (a-c) but with $V = 7$ meV. Black circles indicate Dirac points of top and bottom layers.

## Supplementary Tables

**Supplementary Table 1.**

| LATBG device | Onset, $B^*$ (mT) | $\delta n = \frac{4eB}{h}$ (cm$^{-2}$) | Modelled $\delta E$ (meV) | Modelled $\delta n$ (cm$^{-2}$) | Actual $\delta n$ (cm$^2$) |
|---|---|---|---|---|---|
| D=0 V/nm | 100 | $1 \times 10^{10}$ | 4 | $5 \times 10^9$ | $5 \times 10^9$ |
| D=-0.75 V/nm ($\Delta = 2.5$ meV) | 6 | $6 \times 10^8$ | 0.5±0.2 | 3.0±0.2×$10^8$ | NA |

**Supplementary Table 2.** The table shows if the gaps at Dirac points and half-fillings present in the band structure under specific conditions.

|  |  | AB | BA | Small shift |
|---|---|---|---|---|
| *No displacement* | Half-filling | No | Yes | Yes |
|  | Dirac point | No | No | No |
| *With displacement* | Half-filling | Yes | Yes | Yes |
|  | Dirac point | Yes | No | Yes |

**Supplementary Note 1. Calculating layer specific charge density in LATBG and LATTG devices.**

To calculate the on-layer carrier density and the positions of the CNPs, we employed the model from ref.[1], which accounts for the electrostatic screening of the applied electric field by the charged graphene layers.

In short, each graphene has an effective out-of-plane polarizability, which yields a correction to potential difference between two layers in external displacement field:

$$U_t - U_b = \frac{ed}{\varepsilon_0 \varepsilon}\left(\varepsilon_0 D - \frac{\varepsilon+1}{4}e(n_t - n_b)\right), \tag{1}$$

where $U_{t/b}$, $n_{t/b}$ are on-layer potential energies and on-layer charge densities, $\varepsilon = 2.5$ and $\varepsilon_0$ are an out-of-plane dielectric constant of graphene and vacuum susceptibility, $d = 3.44$ Å is an interlayer distance. This equation gives energetic offset between Dirac cones of top and bottom graphene layers, it can be joined with charge conservation condition and relation between charge density and chemical potential to make a full system of equations:

$$n_{\text{tot}} = n_t + n_b;$$

$$n_{t/b} = \frac{2k_B^2 T^2}{\pi v_F^2 \hbar^2} \cdot \left(Li_2\left[-exp\left(-\frac{\mu_{t/b}}{k_B T}\right)\right] - Li_2\left[-exp\left(\frac{\mu_{t/b}}{k_B T}\right)\right]\right), \tag{2}$$

where $Li_2$ is a dilogarithm function.

These equations allow to recalculate a pair of ($n_{\text{tot}}$, $D$) values to the pair of on-layer charge densities or chemical potentials, which is used to plot Fig. 2d of the main text and Supplementary Fig. 2. A parabolic shape of the Landau fan diagram shown in Supplementary Fig. 2a becomes linear in Supplementary Fig. 2b, c after replotting it with the layer specific carrier density. We note that previous works[2–4] have attributed Landau level bending observed in Landau fan diagrams to the interplay between coexisting high- and low DoS bands. Our calculation of the layer-resolved carrier density confirms that a similar mechanism is at play in our devices. We also note, that in the limit of low temperature, analytical expression for the charge neutrality of individual layers which is used to find the positions of CNPs in the main text is

$$\pm D = \frac{\hbar v_F \varepsilon \sqrt{\pi n_{\text{tot}}} \text{sign } n_{\text{tot}}}{ed} + \frac{e n_{\text{tot}}(\varepsilon+1)}{4\varepsilon_0}. \tag{3}$$

This model can be also extended to the LATTG case, where interlayer potential energy differences are

$$U_3 + \delta_3 - U_1 - \delta_1 = \frac{ed}{\varepsilon_0 \varepsilon}\left(2D\varepsilon_0 - \frac{\varepsilon+3}{4}e(n_1 - n_3)\right); \tag{4}$$

$$U_2 + \delta_2 - U_1 - \delta_1 = \frac{ed}{\varepsilon_0 \varepsilon}\left(D\varepsilon_0 - \frac{\varepsilon+1}{4}e(n_1 - n_2) + \frac{1}{2}en_3\right). \tag{5}$$

In the equations above, we introduced energy shifts $\delta_i$ to account for small doping of graphene layers due to proximity with neighbouring crystals. These shifts are responsible for symmetry breaking on ($n_{\text{tot}}$, $D$) maps shown in the main text, and are determined from the measured positions of CNPs at zero displacement field.

**Supplementary Note 2. Additional characterization of the band gap observed at the CNP of graphene.**

In this Supplementary Note, we present additional measurements of LATBG device demonstrating the properties of the observed gap from Fig. 2 of the main text. In Supplementary Fig. 3a, we show dual gate map of longitudinal resistance measured at $B$=0.7T, when both layers are quantized. The checkered structure originates from intersecting LLs of both layers, a transcription of LLs is shown in Supplementary Fig. 3c, where LLs of each layer are labeled. We further employ measurement geometry from Ref[5] allowing us to detect current going through the bulk of the sample by cutting out contributions of edge channels near the collector terminal (see inset of Supplementary Fig. 3c). Corresponding measurements are shown in Supplementary Fig. 3b, where the same pattern of LLs is observed. In this figure, suppressed bulk current is indicative of a band gap; note that in suspected gaps at CNPs of each layer, the bulk current is suppressed. In Supplementary Fig. 3b, the gap is observed for $D$>0.05 V/nm, access to lower displacement fields is obstructed by artifacts originating from p-n junctions at the voltage probes.

Evidence of the band gap was found in Supplementary Fig. 2c, where derivative of magnetoresistance is plotted as a function of charge density in bottom layer, gap boundaries in the top layer are shown by vertical dashed lines, and magnetic focusing lines are shown by black arrows. Note that magnetic focusing lines start at the boundaries of the gap, suggesting zero DoS in corresponding layer at CNP.

We further confirm this claim by analyzing the fan diagram of bulk current shown in Supplementary Fig. 4. By plotting its derivative, we observed that graphene has suppressed conductivity when the Fermi level is tuned into the gap, and this behavior persists even at zero magnetic field.

Next, we investigated the dependence of the gap on displacement field $D$. For that we measured dual gate map at $B$=100 mT (see Supplementary Fig. 5a) and plotted it as a function of chemical potential in bottom layer and $D$, Supplementary Fig. 5b. Transcription of observed LLs is shown in Supplementary Fig. 5c, where one can see that positions of resistivity peaks corresponding to LLs of the bottom layer are independent of $D$, including the boundaries of the gap. In Supplementary Fig. 5c, gap boundaries in the top layer are traced by two parallel tilted lines, implying that gap size in the top layer does not depend on $D$ as well.

Finally, we measure fan diagrams under applied $D$ at different temperatures, see Supplementary Fig. 6. As one can see, resistivity features corresponding to LLs and gap boundaries persist up to 15K. At the same time, the gap size remains unaffected by temperature.

Overall, from our measurements we can conclude that the gaps at CNPs of graphene layers in LATBG are independent of $D$ (screening) or temperature.

**Supplementary Note 3. Estimation of transport electron mobility in our devices.**

Our experimental geometry with graphene layers stacked directly on top of each other does not allow us to extract transport mobility of individual layers under applied screening. To overcome this limitation below we developed a toy model to estimate the mobility of graphene layers in our devices.

Clean graphene samples exhibit ballistic transport at elevated doping; in this case, formulas, usually derived from Boltzmann equations, become ill-defined. Instead, conductivity should be written using Landauer-Buttiker formalism:

$$\sigma = \frac{2e^2}{h} k_F v_F \tau, \tag{6}$$

where $k_F$ is the Fermi wave vector, $v_F \approx 10^6 \, \frac{m}{s}$ is the Fermi velocity of graphene, $\tau$ is the scattering time, $h$ is the Planck constant, $e$ is the absolute value of electron charge.

In the literature, electron mobility is usually estimated from Drude formula, which, if we take Landauer-Buttiker conductivity, yields:

$$\mu = \frac{\sigma}{ne} \propto \frac{1}{\sqrt{n}}. \tag{7}$$

This quantity diverges at CNP in the absence of disorder. However, if e-h puddles are included in the picture, ballistic transport transitions to diffusive with mean free path $l_{mfp}$ limited by remnant charge density $\delta n$.

For simplicity we estimate average mean free path as $l_{mfp} = \beta/\sqrt{\delta n}$, where $\beta$ is a numerical coefficient responsible for geometry of the scatterer (further set to unity for simplicity).

This estimate is applicable for scattering on point-like defects with sharp potential, which is valid in our case, considering the size of e-h puddles to be of a few tens of nm, and that the long-range part of the Coulomb potential in LATBG is suppressed.

State-of-the art graphene devices are intrinsically pure and suffer mostly from the disorder manifested as charge or energy inhomogeneity. Below, we first derive a model for diffusive transport in ideal graphene and then introduce disorder. Mobility of pristine graphene at finite temperature is given by[6]:

$$\mu(E_F) = e \int dp_x dp_y \tau(\mathbf{p}) f'(E(\mathbf{p}) - E_F) m^{-1}(\mathbf{p}) / \int dp_x dp_y f'(E(\mathbf{p}) - E_F). \tag{8}$$

Here $f(E - E_F) = 1/\left[1 + \exp\frac{E - E_F}{k_B T}\right]$ is the Fermi-Dirac distribution with $k_B$ being the Boltzmann constant, $m^{-1}(\mathbf{p})$ is the inverse mass tensor.

We consider the following dispersion of gapped graphene, observed in Figs. 1, 2 of the main text:

$$E(\mathbf{k}) = \sqrt{\hbar^2 v_F^2 k^2 + \Delta^2}, \tag{9}$$

where $\hbar$ is the reduced Planck constant, and $2\Delta$ is the band gap. In the limit of zero gap $\Delta \to 0$, it becomes a linear spectrum of graphene. We can further substitute scattering time as:

$$\tau = \frac{l_{mfp}}{v_{group}} = \frac{l_{mfp}\sqrt{\hbar^2 v_F^2 k^2 + \Delta^2}}{\hbar v_F^2 k}. \tag{10}$$

As a result, expression (8) gives mobility as a function of Fermi level position and mean free path $\mu = \mu(E_F, l_{mfp})$.

Next, to account for the presence of electron-hole puddles we note that graphene with charge inhomogeneity can be viewed as pristine graphene with spatially non-uniform doping. According to [7], the chance to find a particular charge density at an arbitrary point of the sample obeys normal distribution, thus we can write average mobility as:

$$\mu(n(E_F), \delta n) = \frac{\bar{\sigma}}{ne} = \sum \frac{\mu_i n_i e}{ne} = \int \mu(\rho) \times \frac{\exp -\frac{(\rho-n)^2}{2\delta n^2}}{\sqrt{2\pi \delta n^2}} d\rho. \tag{11}$$

Here we assumed that conductivity of the whole graphene is a weighted sum of conductivities of its differently charged regions.

Further we should consider that mean free path depends on doping: charged graphene is ballistic with $l_{mfp} = w$, where $w$ is the minimal size of a rectangular Hall bar device, whereas in vicinity of CNP scattering is determined by electron hole puddles giving $l_{mfp} = 1/\sqrt{\delta n}$. In the calculations, we smoothly join these two regimes by introducing

$$l_{mfp}(n) = \frac{1}{\sqrt{\delta n}} + \left(w - \frac{1}{\sqrt{\delta n}}\right) \times \frac{1}{1 + \exp \gamma(|n| - \delta n)}, \tag{12}$$

where $\gamma$ is a numerical factor responsible for the sharpness of transition between ballistic and diffusive regimes. In our further modelling we assumed it as $\gamma = -0.5 \delta n$.

Supplementary Fig. 7 shows mobility for different gap sizes and levels of disorder calculated within developed toy model. In Supplementary Fig. 7a we assumed ballistic transport over 5 um scale, while setting the inhomogeneity level to 0. This plot shows the top boundary of the possible graphene mobility in our experiment. In Supplementary Fig. 7b we set inhomogeneity level as $\delta n = 3 \times 10^8 cm^{-2}$ which corresponds to the estimated inhomogeneity of our devices (see estimations in Supplementary Note 4). Under such condition the mobility reaches 16 × 10⁶ cm²/Vs and becomes comparable with the state-of-the-art GaAs 2DEGs.

The uncertainty of the quantization onset value (4.5-6.5 mT for LATBG device) leads to uncertainty of δn, giving 2.1-3.6×10⁸ cm⁻² range, which corresponds to 12-20×10⁶ cm²/Vs range of electron mobility. Furthermore, the resolution of quantization onset is limited by the contact width, and we are using the condition of the 1st LL resolution which is stronger than 1st cyclotron gap resolution condition, thus, the real mobility of the screened graphene devices is likely even higher, and the obtained values should be treated as the bottom boundary of mobility in our devices.

We also note that the mobility estimation provided by this model is sensitive to both, the gap size, and to the inhomogeneity level. For the completeness, in Supplementary Fig. 7b, we have plotted the mobility for the gapless graphene spectrum, for two different inhomogeneity levels. As expected, the higher inhomogeneity level yields lower mobility, and the absence of the gap also leads to the lower mobility for the given value of the mean free path.

**Supplementary Note 4. Onset of Landau quantization in single layer graphene devices.**

In real samples, finite disorder produces spatial fluctuations of charge $\delta n$ which result in smearing of Fermi level $\delta E$. In magnetoresistance measurements, the gap sizes between Landau Levels (LLs) of quantized graphene continuously change with magnetic field; the quantization is resolvable when the gap sizes are big enough to withstand smearing due to disorder, making it a probe for disorder level.

To link the onset of Landau quantization to $\delta n$ and $\delta E$, we construct averaged density of states (DoS) function using similar considerations as in the model of electron mobility. We start with the expression of DoS of quantized graphene at finite temperature adapted from Ref[7] for the gapped spectrum:

$$DoS = \frac{4eB}{h\, k_B T}\left(\sum_{N=-\Lambda, N\neq 0}^{N=\Lambda} Sech^2\left[\frac{(E_{LL}(N)-\mu)}{k_B T}\right] + \frac{1}{2}Sech^2\left[\frac{(\Delta-\mu)}{k_B T}\right] + \frac{1}{2}Sech^2\left[\frac{(-\Delta-\mu)}{k_B T}\right]\right). \tag{13}$$

Here $B$ is the magnetic field, $\mu$ is the chemical potential, $E_{LL}(N,B,\Delta) = \sqrt{2v_F^2 \hbar eB|N| + \Delta^2}\, sign(N)$ is the energy of $N^{th}$ LL, $Sech$ is a hyperbolic secant function.

The average DoS of the disordered sample near Fermi level is given by

$$\overline{DoS}(E_F) = \int_{-\infty}^{\infty} d\mu\, DoS(\mu) \frac{\exp-\frac{(\mu-E_F)^2}{2\delta E^2}}{\sqrt{2\pi}\delta E}. \tag{14}$$

By plotting the averaged DoS map as a function of $\delta E$ and $E_F$, we can determine amplitude of energy fluctuations from the resolution of LLs (see Supplementary Fig. 9). To convert $\delta E$ to $\delta n$, let us notice that the actual charge density of the sample can be expressed both ways:

$$\langle n \rangle = \langle n \rangle_E = \langle n \rangle_n = \sqrt{n^2 + \delta n^2}, \tag{15}$$

which gives the relation

$$\delta n = \sqrt{\langle n \rangle_E^2 - n^2}. \tag{16}$$

Note that the usual way of calculating inhomogeneity from the onset of quantization

$$\delta n = 4eB/h \tag{17}$$

gives overestimated inhomogeneity at zero $D$ (Table S1). On the other hand, $\delta n$ calculated as $\langle n \rangle_E$ for $\delta E$ obtained from DoS map gives an actual $\delta n$ (see Supplementary Fig. 9b and Table 1).

At base temperature and $B$=6 mT, the first cyclotron gap in quantized graphene with $\Delta = 2.5$ meV can be resolved at $\delta E \approx 0.5$ meV, which corresponds to $\delta n = 3 \times 10^8$ cm$^{-2}$ (see Supplementary Fig. 9a and Table 1). Notably, the size of the first cyclotron gap is $\Delta E = 1.3$ meV, which is 2-3 times bigger than $\delta E$ required to close it. In contrast to LATBG, state-of-the-art hBN-encapsulated devices quality is limited by $\delta E = 3$ meV ($\delta n = 2.4 \times 10^9$ cm$^{-2}$ for a typical onset of 50 mT[8]).

Finally, we would like to discuss another factor limiting the resolution of Landau quantization, which is the effect of contact width $W$. Qualitatively, to resolve the onset of the quantum Hall effect, the width of the voltage probe $W$ must be larger than the size of the cyclotron orbits $2R_c$, which would correspond to the filling factor 2 for the largest cyclotron gap at the given magnetic field. This can be written as $2R_c = \left(\frac{\nu\Phi_0}{\pi B}\right)^{1/2} < W$, where $\Phi_0$ is the flux quantum. At 6 mT, this expression gives W~0.7 μm, which coincides with the probe width measured in AFM, possibly limiting the resolution of the quantization at small magnetic fields in our samples.

**Supplementary Note 5. Determination of quantization onset.**

As demonstrated in the Supplementary Note 4, the onset of Landau quantization is directly associated with the homogeneity of the sample. Here, we analyse each measured sample to determine the quantization onset. It is important to clarify our definition of quantization onset. In monolayer graphene, as the external magnetic field increases, Landau levels (LLs) form, leading to DoS modulation and, consequently, resistivity modulation, usually referred to as Shubnikov–de Haas oscillations (SdHO). At higher fields, the system enters the quantum Hall regime, where the gaps between LLs become sufficiently large to prevent the overlapping of broadened LLs. This regime is typically characterized by vanishing $\rho_{xx}$ and quantized plateaus in $\rho_{xy}$.

In the low magnetic field range (below 100 mT), shown in Figs. 1e, 2b, 3e, and 3g of the main text, at the given carrier density, only one graphene layer undergoes quantization, while the other, more doped layers remain unquantized. As a result, in the low magnetic field regime, a significant portion of the current flows through the doped unquantized layers. Consequently, resistance modulations corresponding to the SdHO of the undoped layer appear on the background dominated by ballistic and mesoscopic transport in the other layers.

To precisely determine the SdHO onset and enhance the visibility of oscillations, we subtract the background signal from the raw data as $\Delta R = R - \langle R \rangle$, where $\langle R \rangle$ represents the fast Fourier transform (FFT) smoothed data over a wide data-point window (see Supplementary Fig. 10). The results are shown in Supplementary Fig. 11 for the LATBG device, and in Supplementary Fig. 12 for the LATTG devices.

To establish the quantization onset, we analyse resistance curves $\Delta R(\mu)$ at fixed magnetic field values, as shown in Supplementary Fig. 11a. By comparing the colormap of $\Delta R$ (Supplementary Fig. 11 b) and individual resistance curves, we correlate peaks or dips with the expected positions of the first few LLs and corresponding energy gaps.

In Supplementary Fig. 11a, clear resistance peaks are observed at 30 mT, coinciding with the expected positions of the lowest LLs. As the magnetic field decreases, the peaks weaken and shift in accordance with the calculated LL positions. The evolution of the peak corresponding to N=-1 in the top layer can be continuously tracked down to 6 mT, where the peak remains higher than background signal, establishing a robust upper boundary for the onset of SdHO in this device. It is also possible to extend this boundary down to 5 mT, where a resistance peak remains visible at the expected energy of the N=-1 LL. However, at this field, the peak amplitude is comparable to background fluctuations, making its identification less reliable.

In addition to the resistivity peaks marking LL positions, we can also trace dips corresponding to the energy gaps between the lowest LLs, as shown in Supplementary Fig. 11. By comparing the resistance curves, we track the gap between the and LLs (-2|-1 and $0^+$|1 gaps) down to 8 mT in the bottom layer and 6 mT in the top layer (-2|-1 gap).

To accurately establish the onset of LL formation, it is important to determine the offset from zero magnetic field caused by residual flux trapped inside the superconducting solenoid used to generate the magnetic field in our experiments. The data presented in the main text and in Supplementary Figures already accounts for this offset.

For the LATBG device (e.g. Supplementary Fig. 11), we determine the remnant magnetic field $\delta B$ by fitting LL peaks using

$$B = \frac{(\mu^2 - \Delta^2)}{2\hbar e v_F^2 N} + \delta B \tag{18}$$

with an uncertainty of $\pm 0.5$ mT. Taking the previously determined 5–6 mT limit for the onset of SdHO and accounting for zero-field determination errors, we obtain an SdHO onset range of 4.5-6.5 mT in the LATBG device.

In Supplementary Fig. 12, we perform the same analysis for LATTG devices A and B, shown in Fig. 3 of the main text. In both cases, the remnant magnetic field is obtained as the position of the symmetry axis of the LL parabolic pattern, particularly well visible for device B (see Supplementary Fig. 11d). The uncertainty in this procedure is dictated by the 1 mT magnetic field step.

For device A, the LL peak magnitudes become comparable to the residual background below 7 mT, and the lowest field at which we resolve resistance peaks at LL positions is 5 mT, yielding 5-7 mT range on the colormap coordinates and 4.5-7.5 mT in terms of actual magnetic field value. For device B, we analyse the positive magnetic field curves, where the N=±1 and N=2 peaks are well resolved at 6 mT, with additional less pronounced peaks visible at 5 mT. While this formally suggests 5-6 mT range on the map and 4.5-6.5 mT range in terms of real magnetic field, the resolution of the second LL requires higher sample quality, indicating that the onset is more likely near the lower boundary of this range.

So far, we have discussed only the onset of SdHO. For a complete picture, it is useful to assess the onset of the quantum Hall regime, characterized by well-resolved energy gaps and the formation of robust edge states. Unlike conventional graphene devices, where the quantum Hall onset is typically identified by observing vanishing $\rho_{xx}$ and quantized plateaus in $\rho_{xy}$, in our twisted devices at small magnetic fields, these signatures are smeared by current propagation through the highly doped, non-quantized layer, and while layers are well decoupled in the bulk, they are connected at the device edges.

Although we cannot distinguish the quantum Hall onset using resistivity features, we can characterize the overlap of LL bands via the spectroscopic capabilities of our LATBG device, which should provide an equivalent criterion. LL bands smearing results in the LL resistance peak broadening. We can certainly determine the position $t$ of the LL centre (as a local resistance maximum) and the position of the energy gap (which coincides with a local resistance minimum), then we can find the band broadening $w$ as a FWHM of the peak (see Supplementary Fig. 10). For the top graphene layer at 18 mT it is $w_{-1} = 1.2$ meV, and $w_{0-} = 1.6$ meV. To define the gap resolution, we assume a gaussian shape of the broadened LLs, then we can quantify the resolution via a gaussian peak resolution parameter:

$$R = 1.18 \frac{t_1 - t_2}{(w_1 + w_2)} = 1 \tag{19}$$

where $t_{1,2}$ are peak positions, $w_{1,2}$ are peak FWHM values. This condition corresponds to a 2.3% overlap between the two peaks for $R = 1$. Since the LL fit describes the data well, we apply the condition analytically for the gap between 0th and 1st LLs:

$$B = \left[\frac{(w_{\pm 1} + w_{0\pm})^2}{1.39} R^2 + 2.36 R \Delta (w_{\pm 1} + w_{0\pm})\right] \frac{1}{2\hbar e v_F^2} \tag{20}$$

This gives 13 mT quantum Hall effect onset for R=1, $\Delta = 2.5$ meV (2Δ is the CNP gap) and for $w$ values calculated above. Let us note that visually the width of LLs does not change in the studied region of magnetic fields (Supplementary Fig. 11 b), however it is predicted that LL band broadening has square root dependence on the magnetic field[9] thus 13mT is the upper boundary for the quantum Hall regime onset.

In case of LATTG devices, it is challenging to accurately recalculate $D$ to the chemical potential, however, judged visually, the ±1 and zeroth LLs in Supplementary Fig. 12 are well separated even below 10 mT.

**Supplementary Note 6. Charge inhomogeneity screening model.**

In this Supplementary Note, we describe calculation of charge inhomogeneity in a graphene layer induced by external impurities in the presence of the second graphene layer. It is assumed that one can control the carrier density in the second layer, in which case it can act as a screening substrate to screen the electric field created by the impurities. In what follows, we will refer to the two graphene layers by their labels 1 and 2, respectively.

We distribute the impurities with number density $n_{\text{imp}} = 10^{10}$ cm$^{-2}$ randomly with uniform probability density in a plane at a distance from the first graphene layer. To calculate the density inhomogeneity, we will use self-consistent Hartree approximation (thus neglecting exchange effects). Further assuming that the density significantly varies only on a scale larger than the local Fermi wavelength, we will settle for the semiclassical Hartree approximation, i.e., Thomas—Fermi theory.

The pair of equations describing the number densities in the two layers, $n_1$ and $n_2$, satisfy the following system of equations

$$\mu_1(\boldsymbol{r}) + \frac{e^2}{\kappa}\sum_i \frac{q_i}{\sqrt{(r-r_i)^2+z^2}} + \frac{e^2}{\kappa}\int d\boldsymbol{r}' \frac{n_2(r')-\bar{n}_2}{\sqrt{(r-r')^2+d^2}} + \frac{e^2}{\kappa}\int d\boldsymbol{r}' \frac{n_1(r')}{|r-r'|} = C_1; \qquad (21)$$

$$\mu_2(\boldsymbol{r}) + \frac{e^2}{\kappa}\sum_i \frac{q_i}{\sqrt{(r-r_i)^2+(z-d)^2}} + \frac{e^2}{\kappa}\int d\boldsymbol{r}' \frac{n_2(r')-\bar{n}_2}{|r-r'|} + \frac{e^2}{\kappa}\int d\boldsymbol{r}' \frac{n_1(r')}{\sqrt{(r-r')^2+d^2}} = C_2, \qquad (22)$$

where $q_i$ are the impurity charges in units of the electron charge, which take values of $+1$ or $-1$ randomly with equal probability, $\boldsymbol{r}_i$ are their in-plane positions picked at random with uniform probability density, $\kappa = 4$ is the hBN dielectric constant, $d$ is the distance between two graphene layers, $z$ is the distance between the first graphene layer and the plane of impurities, $\bar{n}_2$ is the average electron number density in the second graphene layer (while the first layer is always left neutral). The local chemical potentials $\mu_1$ and $\mu_2$ that also appear above are related to electron number densities as follows

$$n_1(\boldsymbol{r}) = \frac{\mu_1(r)|\mu_1(r)|}{\pi v_F^2 \hbar^2}; \qquad (23)$$

$$n_2(\boldsymbol{r}) = \frac{\mu_2(r)|\mu_2(r)|}{\pi v_F^2 \hbar^2}, \qquad (24)$$

where $v_F = 10^6$ m/s is the Fermi velocity of electrons in graphene. Constants $C_1$ and $C_2$ are picked in such a way as to ensure that the densities average to 0 and $\bar{n}_2$ in layers 1 and 2, respectively.

Equations (21)— (24) are solved by iterations: On the zeroth iteration, $\mu_1$ and $\mu_2$ are expressed from Eqs. (21) and (22) under the assumption of no fluctuations, with $C_1$ and $C_2$ determined as above, and substituted into Eqs. (23) and (24). This gives $n_1$ and $n_2$, which are then substituted back into Eqs. (21) and (22), where the procedure again repeats itself. To ensure convergence, before each iteration the starting value of the electron densities should be taken as a weighted average of the old one and the one obtained on the previous iteration from Eqs. (23)— (24).

After the charge fluctuations in layer 1 are obtained to the desired accuracy, the level of the inhomogeneity is estimated as $\delta n_1 = \sqrt{\int d\boldsymbol{r}\, n_1^2(\boldsymbol{r})/A}$, with $A$ the area of the graphene layer.

We have investigated the dependence of inhomogeneity on the distance $z$ between layer 1 and impurities and on density in screening layer 2 with inter graphene layer distance fixed at $d = 0.33\ nm$ (Figs. S13 and S14) and on the inter graphene layer distance $d$ with the distance between layer 1 and impurities fixed at $|z| = 20\ nm$, which is approximately half the thickness of hBN layers used in our devices (Supplementary Fig. 15).

Supplementary Fig. 13 shows the size of inhomogeneities for 21 different disorder realizations. One can see that for $z$ above 4 nm an impurity realization of $n_{\mathrm{imp}} = 10^{10}\mathrm{cm}^{-2}$ distributed with uniform probability density can induce an inhomogeneity of a few fractions of $10^{10}\ cm^{-2}$ at charge neutrality (Panel (a)). This will then be reduced by a factor of 3—6 for $z$ in the range 4—10 nm and by a factor of 5—16 for $z$ above 10 nm if layer 2 is charged to $\bar{n}_2 = 0.5 \cdot 10^{12}\mathrm{cm}^{-2}$ (Panels (b) and, primarily, (c)). The inhomogeneities at charge neutrality and doping averaged over the disorder realizations are shown in Supplementary Fig. 14, Panel (a). Their ratio is shown in Supplementary Fig. 14, Panel (c).

We also investigated the dependence of density inhomogeneity on the distance between graphene layers, Supplementary Fig. 15. It shows that the presence of a charge neutral second graphene layer only slightly reduces the size of density fluctuations, as the distance changes from 30 nm to a few angstroms, Supplementary Fig. 15a. When the second layer is charged to $\bar{n}_2 = 0.5 \cdot 10^{12}\mathrm{cm}^{-2}$ (Panels (b) and (c)), the reduction of inhomogeneities is as low as a factor of 1—2 for $d$ above 10 nm. As $d$ decreases to a few angstroms the reduction quickly increases to a factor of 5—13.

To further illustrate the importance of proximity in screening density fluctuations we calculated the density distribution in layer 1 with one disorder realization at $d = 5\ \mathrm{nm}$ and $d = 0.33\ \mathrm{nm}$, Supplementary Fig. 16. When layer 2 is charged to $\bar{n}_2 = 0.5 \cdot 10^{12}\mathrm{cm}^{-2}$, reduction of inhomogeneities at inter graphene layer distance of 0.33 nm (Panel (c)) appears more significant than at 5 nm (Panel (b)).

Finally, we would like to show that the screening strength does not depend on which side of the LATBG the impurities are located (or equivalently, does not depend on the choice of the screening layer). In Supplementary Fig. 17, we calculate inhomogeneity when the impurities and the screening layer are on the opposite sides as opposed to Supplementary Fig. 13. Both cases provide the same screening strength.

## Supplementary Note 7. LATBG fan diagram modelling.

We consider a model of a 4-terminal Hall bar made of weakly-coupled graphene layers with weak current tunnelling between them (see Supplementary Fig. 18). Such an assumption is consistent with the earlier experiments, which point out relatively large interface resistance in such a system. In this situation, the contacts play a special role by providing additional conductivity between the two layers. The strength of contact-induced interlayer conductivity significantly influences the predicted measurement results.

To model the fan diagram shown in Fig. 2b of the main text, below we consider the following situation. One of graphene layers is highly doped, so we can neglect the quantisation in this layer and assume that it has a large classical conductivity $\sigma_{xx1} \geq \sigma_{xy1}$. At the same time, the second layer is doped closer to charge neutrality point, and is in the Quantum Hall regime, $\sigma_{xx2} \leq \sigma_{xy2}$, while $\sigma_{xx2}, \sigma_{xy2} \ll \sigma_{xx1}$. As an example, we assume Gauss-broadened Landau levels at $\sigma_{xx2}$, and similar Gauss broadened $\sigma_{xy2} = \frac{e^2}{h}\nu$, where $\nu$ is Landau level filling factor. Additionally, we assume the presence of the gap at the neutrality point of quantised layer, which we model as $\sigma_{xx2} = \sigma_{xy2} = 0$, as shown in Fig S19.

The results of the measurements are simulated using a numerical solution of the following system of equations:

$$\vec{j}_{1,2} = \vec{\sigma} \cdot \vec{\nabla} U_{1,2} ; \tag{25}$$

$$\vec{\nabla} \cdot \vec{j_1} = s(x)(U_2 - U_1) ; \tag{26}$$

$$\vec{\nabla} \cdot \vec{j_2} = s(x)(U_1 - U_2), \tag{27}$$

where $s(x)$ is the interlayer conductivity per unit area, which can receive contributions from the bulk, the edge and the contact area of the sample, $U_{1(2)}$ is the potential of the 1(2) layer, $\vec{\sigma}$ is the conductivity tensor, $j_{1(2)}$ is the current in 1(2) layer.

To understand the results, let us start by analysing a few different limits. First, if the interlayer tunnelling is large, the conductivity tensors of both layers are summed up and can be easily inverted to give

$$R_{xx} \xrightarrow{s \to \infty} \frac{\sigma_{xx1} + \sigma_{xx2}}{(\sigma_{xx1} + \sigma_{xx2})^2 + (\sigma_{xy1} + \sigma_{xy2})^2}. \tag{28}$$

This formula results in small dips in $R_{xx}$ at the location of Landau levels due to dominance of $\sigma_{xx1}$, this is not what was observed experimentally. Another limit is $s \to 0$, but here we need a careful consideration of the voltage-probe contacts: If we assume ideally-connected voltage contacts, these would shunt the two layers in the contact region, the voltage at the side-contacts will be mainly determined by layer 1, where a dominant current flows, corrected by a small 2-terminal Hall conductivity of layer 2 ($\approx 1/|\sigma_{xy2}|$) between the voltage probes:

$$R_{xx} \xrightarrow{s \to 0, \ s_c \to \infty} \frac{\sigma_{xx1} + |\sigma_{xy2}|}{(\sigma_{xx1} + |\sigma_{xy2}|)^2 + (\sigma_{xy1})^2} \xrightarrow{\sigma_{xy1} \to 0} \frac{1}{\sigma_{xx1} + |\sigma_{xy2}|}, \tag{29}$$

which would not produce peaks at Landau level locations, because it is almost independent of $\sigma_{xx2}$.

Finally, if the conductivities $\sigma_{c1,2}$ between voltage contacts and layers 1,2 are small enough, a linear combination $V = \frac{\sigma_{c2}V_1 + \sigma_{c1}V_2}{\sigma_{c1} + \sigma_{c2}}$ of layer voltage drops would be measured. The measured resistance is then given by $R_{xx} = V/(I_1 + I_2)$. Taking into account that $I_1 \approx V_{SD}/R_{xx1}$ in classical regime and $I_2 \approx$

$V_{SD} \frac{\sigma_{xx2}^2+\sigma_{xy2}^2}{|\sigma_{xy2}|+a\,\sigma_{xx2}}$ (where $a$ is width to length ratio of the sample, this expression gives a reasonable interpolation between classical and quantum Hall effect limits), and $V_1 = I_1\,R_{xx1}$; $V_2 = I_2\,R_{xx2}$ where $V_{SD}$ is a source-drain voltage on current contacts, we arrive at

$$R_{xx} = \frac{\sigma_{xx1}[(a\,\sigma_{c1}+\sigma_{c2})\sigma_{xx2}+\sigma_{c1}|\sigma_{xy2}|]}{(\sigma_{c1}+\sigma_{c2})[(a\sigma_{xx2}+|\sigma_{xy2}|)(\sigma_{xx1}^2+\sigma_{xy1}^2)+\sigma_{xx1}(\sigma_{xx2}^2+\sigma_{xy2}^2)]}. \tag{30}$$

This expression is plotted with solid red line in Supplementary Fig. 19. Introducing a small tunnelling between the layers requires a numerical solution of eqs (25)-(27), but the results can be fitted with eq (30) where conductivities of layers are slightly mixed $\sigma_1 \to \sigma_1 + \epsilon\sigma_2$. An example of the result is shown in Supplementary Fig. 19 by a dashed line. We see that both results have a behaviour at CNP that does not agree with experiment.

To improve the description, we consider a model where the K and K' valleys in graphene are not equilibrated. Indeed, if the two graphene layers are decoupled due to momentum mismatch, the valleys can decouple as well for the same reason of momentum mismatch. Hence, each valley will have its own local potential induced by applied voltage and graphene sheet will be described as two independent sheets coupled only at the contacts. Assuming the same contact conductivity, $\sigma_{c2}/2$, to both valleys, we arrive at a generalization of eq (30):

$$R_{xx} = \frac{2\,\sigma_{c1} + \left[\frac{\sigma_{c2}\,\sigma_{xx2K}}{a\sigma_{xx2K}+|\sigma_{xy2K}|}\right]+[...K\to K']}{2\,(\sigma_{c1}+\sigma_{c2})\left(\sigma_{xx1}+\frac{\sigma_{xy1}^2}{\sigma_{xx1}}+\left[\frac{\sigma_{xx2K}^2+\sigma_{xy2K}^2}{a\sigma_{xx2K}+|\sigma xy2K|}\right]+[...K\to K']\right)}. \tag{31}$$

The equation (31) well reproduces experimental result shown in Fig. 2b of the main text.

The discussion above applies for finite magnetic field, however, experimental observations suggest presence of the gap also at zero magnetic field. Furthermore, at zero magnetic field we sometimes observe dips in resistance inside the gap, which is counterintuitive as gapped graphene is expected to have low DoS. At the same time bulk transport indicates low conductivity of the bulk in the gapped region at zero B. These observations may be governed into hypothesis that even at zero magnetic field there are edge valley currents responsible for the dips in resistivity. This suggestion is coherent with the developed model, but a clearer experimental proof requires much more work and lays beyond this work.

**Supplementary Note 8. Possibility of a many-body gap at CNP of graphene.**

Graphene at half-filling was theoretically studied a long time ago, promising a Mott insulator gap at the CNP of graphene[10,11]. However, the experiments with freestanding graphene have shown renormalization of the Fermi velocity, instead of a gap, which was attributed to the presence of strong long-range Coulomb interactions [12,13]. The condition for the phase transition is determined by $\alpha = \frac{e^2}{\varepsilon^* v_F \hbar}$, an analogue of the fine structure constant in graphene; this parameter controls the strength of electron interactions and strongly depends on screening. The threshold value of α for the phase transition was long debated in the literature without achieving any consensus (see Ref[14] for the summary of predicted critical $\alpha$). A recent study suggests that the phase transition is obstructed by competing long-range and short-range Coulomb interactions[15], where the latter favors the semimetal-Mott insulator transition. This makes LATBG an exciting platform for the realization of Mott insulator state as metallic screening of the charged layer should significantly suppress the long-range part of Coulomb potential, making the phase transition more favorable.

To account for complex structure of our device, we treat a test monolayer graphene as being isolated from the other but exhibit in-plane Coulomb e-e interactions affected by Thomas-Fermi screening.

Screened potential $\tilde{\phi}(\mathbf{q}, z)$ reads

$$\tilde{\phi}(\mathbf{q}, z) = -\frac{2\pi e}{\epsilon^*} \frac{e^{-|z|q}}{q+q_s}, \tag{32}$$

where

$$q_s = \frac{4e^2 k_F}{\epsilon^* \hbar v} \tag{33}$$

is the Thomas-Fermi screening reciprocal radius. Here, $k_F$ can be written in terms of $\bar{n}_e$ (surface charge density on the screening layer) as $k_F = \sqrt{\pi \bar{n}_e}$. Note, that if the electron dispersion was parabolic, then $q_s$ would not depend on $n_e$. Hence, the great tunability of $q_s$ is specific for conical bands in graphene.

The typical densities $n_e$ we use to screen the Coulomb interactions in the intrinsic layer is of the order of $10^{11}$ cm$^{-2}$ that corresponds to $q_s^{-1}$ of a few nm – much larger than the interlayer distance. The in-plane electron-electron interactions are therefore described by $V_q = -e\tilde{\phi}(\mathbf{q}, 0)$ written explicitly as

$$V_q = \frac{2\pi e^2}{\epsilon^*} \frac{1}{q+q_s}. \tag{34}$$

In the limit of undoped screening graphene layer $\bar{n}_e \to 0$ we have $q_s = 0$, and $V_q$ represents bare Coulomb interactions in 2D.

Obviously, the effect of screening is substantial only for electrons with small $q \lesssim q_s$, i.e. having long wavelength. In graphene, those electrons are sitting near conical band intersections. This is the reason why we need intrinsic graphene layer near the doped one to see the effect.

The tight-binding Hamiltonian for electrons in graphene should be used in the low-energy (conical) approximation to be compatible with the screening model considered above. Considering electron spin we employ two copies of the same Hamiltonian as

$$H_{tb\uparrow} = \sum_{\mathbf{k}} \begin{pmatrix} c_{A\mathbf{k}\uparrow}^\dagger & c_{B\mathbf{k}\uparrow}^\dagger \end{pmatrix} \begin{pmatrix} 0 & \epsilon_k e^{-i\zeta_k} \\ \epsilon_k e^{i\zeta_k} & 0 \end{pmatrix} \begin{pmatrix} c_{A\mathbf{k}\uparrow} \\ c_{B\mathbf{k}\uparrow} \end{pmatrix}, \tag{35}$$

and

$$H_{tb\downarrow} = \sum_{\mathbf{k}} \begin{pmatrix} c^\dagger_{A\mathbf{k}\downarrow} & c^\dagger_{B\mathbf{k}\downarrow} \end{pmatrix} \begin{pmatrix} 0 & \epsilon_k e^{-i\zeta_k} \\ \epsilon_k e^{i\zeta_k} & 0 \end{pmatrix} \begin{pmatrix} c_{A\mathbf{k}\downarrow} \\ c_{B\mathbf{k}\downarrow} \end{pmatrix}, \tag{36}$$

where $\tan\zeta_k = k_y/k_x$ is the direction of electron momentum, and $\epsilon_k = \hbar v k$. The index A(B) denotes respectively the sublattice A(B).

We postulate the reduced pairing Hamiltonian given by

$$\begin{aligned} H_{\text{pair}} &= \sum_{\mathbf{k}\mathbf{k}_1} V_{\mathbf{k}\mathbf{k}_1} \left(c^\dagger_{A\mathbf{k}\uparrow} c_{A\mathbf{k}\uparrow} - 1/2\right)\left(c^\dagger_{A\mathbf{k}_1\downarrow} c_{A\mathbf{k}_1\downarrow} - 1/2\right) \\ &+ \sum_{\mathbf{k}\mathbf{k}_1} V_{\mathbf{k}\mathbf{k}_1} \left(c^\dagger_{B\mathbf{k}\uparrow} c_{B\mathbf{k}\uparrow} - 1/2\right)\left(c^\dagger_{B\mathbf{k}_1\downarrow} c_{B\mathbf{k}_1\downarrow} - 1/2\right). \end{aligned} \tag{37}$$

Here, the creation and annihilation operators have the same meaning as in the tight-binding Hamiltonian above. The $-1/2$ are added to keep the electron-hole symmetry upon interactions. The matrix element $V_{\mathbf{k}\mathbf{k}_1}$ is given by equation ([34](#)) with $q = |\mathbf{k} - \mathbf{k}_1|$.

The purpose of the Hamiltonian is to describe density-density interactions between electrons having opposite spins and, in this way, open a gap in the gapless conical bands. The standard approach to this problem is the so-called Hubbard model, where the electron-electron interactions are described in a simple but extreme manner: the Coulomb interaction between electrons is taken to be point-like in real space and, hence, a constant in momentum space leading to $V_{\mathbf{k}\mathbf{k}_1} = V_0$. Thus, taking the limit of strong screening in our model ($q_s \to \infty$) should lead to the Hubbard model results, i.e. transition into an insulating state. However, the formal limit

$$\begin{aligned} \lim_{\bar{n}_e \to \infty} V_q &= \frac{\pi \hbar v}{2\sqrt{\pi \bar{n}_e}} \\ &= 0 \end{aligned} \tag{38}$$

not only makes interactions short-range but also diminishes the interaction strength to zero. We therefore must either modify our Hamiltonian to include intersublattice interaction terms or manually add a certain cut-off to separate the long- and short-range interaction effects.

In what follows, we consider the second option and introduce a momentum cut-off in the gap equation which limits the integration within the small $|\mathbf{k} - \mathbf{k}_1|$ corresponding to the long-range interactions in the real space. Obviously, the separation between the long range and the short range depends on the screening radius, hence, the cut-off will also be determined by $q_s$. The cut-off and gap must be deduced self-consistently.

The mean-field approximation can be formalized as

$$\begin{aligned} c^\dagger_{A,B\mathbf{k}\uparrow} c_{A,B\mathbf{k}\uparrow} &= \langle c^\dagger_{A,B\mathbf{k}\uparrow} c_{A,B\mathbf{k}\uparrow} \rangle + \left(c^\dagger_{A,B\mathbf{k}\uparrow} c_{A,B\mathbf{k}\uparrow} - \langle c^\dagger_{A,B\mathbf{k}\uparrow} c_{A,B\mathbf{k}\uparrow} \rangle\right), \\ c^\dagger_{A,B\mathbf{k}_1\downarrow} c_{A,B\mathbf{k}_1\downarrow} &= \langle c^\dagger_{A,B\mathbf{k}_1\downarrow} c_{A,B\mathbf{k}_1\downarrow} \rangle + \left(c^\dagger_{A,B\mathbf{k}_1\downarrow} c_{A,B\mathbf{k}_1\downarrow} - \langle c^\dagger_{A,B\mathbf{k}_1\downarrow} c_{A,B\mathbf{k}_1\downarrow} \rangle\right), \end{aligned} \tag{39}$$

where the expression in the brackets represents a small parameter. It is convenient to introduce the order parameter given by

$$\begin{aligned} \Delta_{A,B\mathbf{k}\uparrow} &= \sum_{\mathbf{k}_1} V_{\mathbf{k}\mathbf{k}_1} \langle c^\dagger_{A,B\mathbf{k}_1\uparrow} c_{A,B\mathbf{k}_1\uparrow} \rangle, \\ \Delta_{A,B\mathbf{k}\downarrow} &= \sum_{\mathbf{k}_1} V_{\mathbf{k}\mathbf{k}_1} \langle c^\dagger_{A,B\mathbf{k}_1\downarrow} c_{A,B\mathbf{k}_1\downarrow} \rangle, \end{aligned} \tag{40}$$

and Eq. (37) then reads

$$H_{\text{pair}}^{MF} = \sum_{\mathbf{k}} \big[ (\Delta_{A\mathbf{k}\uparrow} - U_{\mathbf{k}}/2) c_{A\mathbf{k}\downarrow}^{\dagger} c_{A\mathbf{k}\downarrow}$$
$$+ (\Delta_{A\mathbf{k}\downarrow} - U_{\mathbf{k}}/2) c_{A\mathbf{k}\uparrow}^{\dagger} c_{A\mathbf{k}\uparrow} + (\Delta_{B\mathbf{k}\uparrow} - U_{\mathbf{k}}/2) c_{B\mathbf{k}\downarrow}^{\dagger} c_{B\mathbf{k}\downarrow}$$
$$+ (\Delta_{B\mathbf{k}\downarrow} - U_{\mathbf{k}}/2) c_{B\mathbf{k}\uparrow}^{\dagger} c_{B\mathbf{k}\uparrow} + U_{\mathbf{k}}/2 \quad (41)$$
$$- \Delta_{A\mathbf{k}\downarrow} \langle c_{A\mathbf{k}\uparrow}^{\dagger} c_{A\mathbf{k}\uparrow} \rangle - \Delta_{B\mathbf{k}\downarrow} \langle c_{B\mathbf{k}\uparrow}^{\dagger} c_{B\mathbf{k}\uparrow} \rangle \big].$$

Here, $U_{\mathbf{k}}$ stands for

$$\sum_{\mathbf{k}_1} V_{\mathbf{k}\mathbf{k}_1} = U_{\mathbf{k}}. \quad (42)$$

Combining Eq. (37) and the tight-binding Hamiltonian (35)- (36) the total mean-field Hamiltonian can be written as

$$H^{MF} = \sum_{\mathbf{k}} \Big[ -\Delta_{A\mathbf{k}\downarrow} \langle c_{A\mathbf{k}\uparrow}^{\dagger} c_{A\mathbf{k}\uparrow} \rangle - \Delta_{B\mathbf{k}\downarrow} \langle c_{B\mathbf{k}\uparrow}^{\dagger} c_{B\mathbf{k}\uparrow} \rangle$$
$$+ \frac{U_{\mathbf{k}}}{2} + (c_{A\mathbf{k}\uparrow}^{\dagger} c_{B\mathbf{k}\uparrow}^{\dagger} c_{A\mathbf{k}\downarrow}^{\dagger} c_{B\mathbf{k}\downarrow}^{\dagger}) H_k^{MF} \begin{pmatrix} c_{A\mathbf{k}\uparrow} \\ c_{B\mathbf{k}\uparrow} \\ c_{A\mathbf{k}\downarrow} \\ c_{B\mathbf{k}\downarrow} \end{pmatrix} \Big], \quad (43)$$

where

$$H_k^{MF} = \begin{pmatrix} \Delta_{A\mathbf{k}\downarrow} - \frac{U_{\mathbf{k}}}{2} & \epsilon_k e^{-i\zeta_k} & 0 & 0 \\ \epsilon_k e^{i\zeta_k} & \Delta_{B\mathbf{k}\downarrow} - \frac{U_{\mathbf{k}}}{2} & 0 & 0 \\ 0 & 0 & \Delta_{A\mathbf{k}\uparrow} - \frac{U_{\mathbf{k}}}{2} & \epsilon_k e^{-i\zeta_k} \\ 0 & 0 & \epsilon_k e^{i\zeta_k} & \Delta_{B\mathbf{k}\uparrow} - \frac{U_{\mathbf{k}}}{2} \end{pmatrix}. \quad (44)$$

Diagonalizing $H_k^{MF}$ allows to calculate antiferromagnetic order parameter $M_{\mathbf{k}}$ defined as a difference of sublattice total magnetizations. In case when it is uniform in the momentum space ($M_{\mathbf{k}} = M_0$), the fundamental gap $2M_0$ at k = 0 can be deduced from

$$M_0 = \frac{1}{2} \sum_{\mathbf{k}} V_{\mathbf{k}} \frac{M_0}{\sqrt{\epsilon_k^2 + M_0^2}} \Big[ 1 - 2n_F \Big( \sqrt{\epsilon_k^2 + M_0^2} \Big) \Big]. \quad (45)$$

The sum in equation (45) diverges without cut-off. Taking into account the valley degeneracy (the spin degree of freedom is already taken into account in the four-band Hamiltonian) the gap equation can be written as

$$M_0 = \alpha \int_0^{\varepsilon_c} d\varepsilon \frac{\varepsilon}{\varepsilon + \varepsilon_s} \frac{M_0}{\sqrt{\varepsilon^2 + M_0^2}} \Big[ 1 - 2n_F \Big( \sqrt{\varepsilon^2 + M_0^2} \Big) \Big], \quad (46)$$

where $\alpha = e^2/(\epsilon^* \hbar v)$, $\varepsilon_s = \hbar v q_s$, and $\varepsilon_c$ is the cut-off to be determined. Note that $M_0$ Can be canceled out.

In the limit $T = 0$ equation (46) takes the form

$$1 = \alpha \int_0^{\varepsilon_c} d\varepsilon \frac{\varepsilon}{\varepsilon+\varepsilon_s} \frac{1}{\sqrt{\varepsilon^2+M_0^2}}$$

$$= \alpha \left\{ \ln\left(\frac{\varepsilon_c + \sqrt{M_0^2+\varepsilon_c^2}}{M_0}\right) + \frac{\varepsilon_s}{\sqrt{M_0^2+\varepsilon_s^2}} \right. \tag{47}$$

$$\left. \times \ln\left[\frac{\varepsilon_s}{\varepsilon_s+\varepsilon_c} \frac{M_0^2 - \varepsilon_c\varepsilon_s + \sqrt{(M_0^2+\varepsilon_s^2)(M_0^2+\varepsilon_c^2)}}{M_0\left(M_0+\sqrt{M_0^2+\varepsilon_s^2}\right)}\right] \right\}.$$

To determine the cut-off we set $M_0 = 0$ and look for a solution of equation (47) at $\varepsilon_c \gg \varepsilon_s$, as the cut-off must be much larger than any other energy scale involved. Hence, equation (47) can be written as

$$1 = \alpha \ln\left(\frac{\varepsilon_c}{\varepsilon_s}\right), \tag{48}$$

and

$$\varepsilon_c = \varepsilon_s \mathrm{e}^{1/\alpha}. \tag{49}$$

The cut-off obviously depends on screening. If the screening is strong (large $q_s$), then the cut-off increases to maintain the overall interaction effect. In this sense, our model is phenomenological. As we have discussed above, without the screening-dependent cut-off the screening would not only make interactions short-range but suppress interactions completely.

Equation (47) can be simplified for small $M_0$ ($M_0 \ll \varepsilon_s \ll \varepsilon_c$) and written as

$$\frac{1}{\alpha} = \ln\left[\frac{2\varepsilon_s\varepsilon_c\left(1+\frac{\varepsilon_c}{2\varepsilon_s}\right)}{(\varepsilon_c+\varepsilon_s)(M_0+\varepsilon_s)}\right]. \tag{50}$$

Solutions of equations (47) and (50) are shown in Supplementary Fig. 21 side by side. It is clear that equation (50) is a good approximation at low $n_e$. Thus, we can use it to estimate $M_0$ as

$$M_0 = \frac{\varepsilon_s}{1+\mathrm{e}^{1/\alpha}}. \tag{51}$$

If we take graphene encapsulated into the h-BN with in-plane $\epsilon^* = 6.85$[16,17], then $\alpha = 0.32$, and $M_0 = 0.042\varepsilon_s$. At $n_e = 10^{11}$ cm$^{-2}$ we have $q_s^{-1} \sim 10$ nm, and $\varepsilon_s = 0.047$ eV. That results in $M_0 = 0.002$ eV.

The model suggests gap sizes comparable to those that were measured. However, the predicted gap behavior contradicts the experimental observations, namely the dependence on $n_e$ (or, equivalently, $D$), Supplementary Fig. 5, and dependence on temperature (Supplementary Fig. 6). Therefore, it is unlikely that the observed gap originates from strong electron-electron interactions.

**Supplementary Note 9. Additional LATTG device.**

In this Supplementary Note, we present measurements on an additional LATTG device C, shown in Supplementary Fig. 22a. Similar to the device shown in Fig. 3 of the main text, we first characterized it at zero magnetic field by measuring resistivity as a function of $n_{tot}$ and $D$, Supplementary Fig. 22a. In this map, we observe features corresponding to the CNPs of each graphene layer, which split under the applied displacement field. Similar to the device discussed in the main text, we find that the CNP of the middle layer shifts to negative $n_{tot}$, indicating doping of the encapsulating layers even at zero $D$. This behavior is further corroborated by Supplementary Fig. 22c, which shows a similar double-gate map measured under an applied magnetic field of 0.7 T. In Supplementary Fig. 22c, we also resolve a gap at the CNP of the top graphene layer.

To further characterize this device, we measured the magnetoresistance under an applied displacement field, Supplementary Fig. 22d. In this map, the Landau fans converge around the expected positions of the CNPs for each graphene layer. However, in this device, resolving the quantization of individual layers at small magnetic fields is challenging. For example, in Supplementary Fig. 22e, the Landau fans of individual layers are obscured by magnetic focusing peaks originating from other heavily doped graphene layers.

**Supplementary Note 10. LATBG bandsturcture calculations.**

Atomic structure of LATBG strongly depends on the twist angle: it is crystalline for a countable set of commensurate angles[18], and is non-crystalline for a continuum of twist angles. In real devices, it is extremely hard to precisely achieve a specific commensurate angle and control relative shift of the layers.

To date, there are methods to calculate band structure of LATBG only for commensurate angles with the focus on $\theta = 21.8°$, which produces the smallest unit cell[19–21]. Calculations strongly depend on the twist angle, relative shift, displacement field and interlayer hopping parameters. Since we focus on the band structure near the Dirac points, we further employ a low energy continuum model.

We consider a bilayer graphene with an initial AA stacking at the origin at the point (0, 0). We then rotate the layer l = 1(2) by an angle $\pm\theta/2$, respectively. Low energy state Hamiltonian for completely decoupled bilayer system expanded near K-points is written as

$$H_0 = \begin{pmatrix} \hbar v_F \sigma_{-\theta/2} \cdot \boldsymbol{k} & 0 \\ 0 & \hbar v_F \sigma_{-\theta/2} \cdot \boldsymbol{k} \end{pmatrix}, \quad (52)$$

where $v_F$ is the Fermi velocity. Hamiltonian (52) produces two twofold degenerate Dirac cones for each layer. Out-of-plane displacement field creates energy offset that can be included as

$$H_0 = \begin{pmatrix} \hbar v_F \sigma_{-\theta/2} \cdot \boldsymbol{k} + u\sigma_0 & 0 \\ 0 & \hbar v_F \sigma_{-\theta/2} \cdot \boldsymbol{k} - u\sigma_0 \end{pmatrix}. \quad (53)$$

The main effect of displacement field is the offset of the top and bottom layer bands proportional to $\pm u$.

Next, we consider weak interlayer coupling following Ref[22] by introducing $T_0(\boldsymbol{d})$:

$$H_0 = \begin{pmatrix} \hbar v_F \sigma_{-\theta/2} \cdot \boldsymbol{k} + u\sigma_0 & T_0(\boldsymbol{d}) \\ T_0^\dagger(\boldsymbol{d}) & \hbar v_F \sigma_{-\theta/2} \cdot \boldsymbol{k} - u\sigma_0 \end{pmatrix}. \quad (54)$$

Here $T_0(\boldsymbol{d})$ depends on several parameters, namely on $w_0$, responsible for interlayer hopping from AB to BA sites, on $w_1$, responsible for interlayer hopping at AA regions, and on vector $\boldsymbol{d}$ which describes shift between the layers and, hence, transition between AA, AB, and BA stacking. We use $w_{1(2)} = 5$ meV obtained from correspondence with tight-binding calculations.

Below we analyse spectrum dependence of LATBG with $\theta \approx 21.8°$ given by (54) on displacement field and interlayer shift vector, identifying several possible cases (see Table 2).

In the spectra, Supplementary Fig. 23, the gaps at Dirac points open at finite displacement field for any initial stacking but BA. Additionally, we observe a larger gap at half-filling. Although, we do not observe signatures of a band gap at half-filling in the experiment, the band structure of TBG twisted at high angle strongly depends on stacking and displacement field.

Comparing theoretical calculations with the experiment, we observe that in LATBG Dirac cones of individual layers preserve their linearity only on intermediate energy scale ($\mu_{SLG} \sim 50$ meV), while at lower energies gaps at Dirac points and total half-filling appear. Discussion of low energy layer hybridization is challenging due to the lack of theoretical methods and inability to determine the twist angle from the transport measurements. However, both theory and measurements point towards low energy band reconstruction calling for further studies.

We note that it is likely, that there should be a set of parameters, e.g. twist angle, shift of the layers, and displacement, which would give a better agreement with our experiment, namely the absence of the gap at half-filling, and small gaps at the CNPs of individual layers, which would be independent on displacement field. However, to date, there are no systematic study of high twist angle graphene structures, which would include those parameters, and such study would be beyond the scope of our work.